\documentclass[linenumber]{aastex631}

\shortauthors{Deng et al.}
\graphicspath{{./}{figures/}}
\usepackage{amsmath}
\usepackage{longtable}
\usepackage{graphicx}
\makeatletter

\newcommand{\Rmnum}[1]{\expandafter\@slowromancap\romannumeral #1@}
\makeatother
\begin{document}
\title{On the dynamical evolution of the asteroid belt in a massive
star-neutron star binary}
\correspondingauthor{Yong-Feng Huang}
\email{hyf@nju.edu.cn}
\author{Chen Deng}
\affiliation{School of Astronomy and Space Science, Nanjing
University, Nanjing 210023, China}

\author{Yong-Feng Huang}
\affiliation{School of Astronomy and Space Science, Nanjing
University, Nanjing 210023, China}
\affiliation{Key Laboratory of Modern Astronomy and Astrophysics
(Nanjing University), Ministry of Education, Nanjing 210023, China}

\author{Chen Du}
\affiliation{School of Astronomy and Space Science, Nanjing
University, Nanjing 210023, China}

\author{Pei Wang}
\affiliation{CAS Key Laboratory of FAST, NAOC, Chinese
Academy of Sciences, Beijing 100101, China}
\affiliation{Institute for Frontiers in Astronomy and Astrophysics,
Beijing Normal University, Beijing 102206, China}

\author{Zi-Gao Dai}
\affiliation{Department of Astronomy, School of Physical Sciences,
    University of Science and Technology of China, Hefei 230026, China}

\begin{abstract}

Some fast radio bursts (FRBs) exhibit repetitive behaviors and
their origins remain enigmatic. It has been argued that repeating
FRBs could be produced by the interaction between a neutron star
and an asteroid belt. Here we consider the systems in which an
asteroid belt dwells around a massive star, while a neutron star,
as a companion of the massive star, interacts with the belt
through gravitational force. Various orbital configurations are
assumed for the system. Direct $N$-body simulations are performed
to investigate the dynamical evolution of the asteroids belt. It
is found that a larger orbital eccentricity of the neutron star
will destroy the belt more quickly, with a large number of
asteroids being scattered out of the system. A low
inclination not only suppresses the collisions but also inhibits
the ejection rate at early stages. However, highly inclined
systems may undergo strong oscillations, resulting in the
Kozai--Lidov instabilities. Among the various configurations, a
clear periodicity is observed in the collision events for the case
with an orbital eccentricity of 0.7 and mutual inclination of
$0^{\circ}$. It is found that such a periodicity can be sustained
for at least 8 neutron star orbital periods, supporting this
mechanism as a possible explanation for periodically repeating
FRBs. Our studies also suggest that the active stage of these
kinds of FRB sources should be limited, since the asteroid belt
would finally be destroyed by the neutron star after multiple
passages.

\end{abstract}


\section{Introduction}
\label{sec:intro}

Fast radio bursts (FRBs), millisecond-duration radio
flashes that randomly occur in the sky, have remained
mysterious and become a hot topic in astrophysics since
they were discovered in 2007 \citep{Lorimer2007science,
Thornton2013Science,Spitler2016nature,Bannister2019science,
CHIME2020Natur,xu2022Nature}. The first repeating source
FRB 20121102A was found to be in association with a dwarf
star-forming galaxy at z = 0.19, which confirms the
cosmological origin of
FRBs \citep{Spitler2016nature,Marcote2017ApJ,Tendulkar2017ApJ}.
So far, nearly 800 FRB sources have been detected, of which
more than 60 FRBs are repeaters \citep{CHIME2019ApJ,
CHIME2021ApJS,Kumar2019ApJ,CHIME2023ApJ,Hu2023ApJS}.
Interestingly, FRB 20121102A seems to have a $\sim$ 160-day periodicity
in its repeating activity, and a 16.35-day periodicity is also
found for FRB 20180916B
\citep{Chime2020Natur.582..351C,Rajwade2020MNRAS,Cruces2021MNRAS} .
While there are noticeable differences in the time-frequency
structure, energy spectrum, and polarization characteristics
between one-off and repeating FRBs \citep{CHIME2019ApJ,
CHIME2021ApJS,Pleunis2021ApJ,Zhou2022RAA,Chen2022ApJ},
the classification of FRBs based on observations are still challenging
\citep{Luo2020Natur,xu2022Nature}. The variations in
the observed Rotation Measure (RM) and even RM revesal
indicate that the central engine of FRBs may reside in
a complex magnetized plasma environment
\citep{Michilli2018Natur,xu2022Nature,Niu2022Natur,
Wang2022NatCo,Anna-Thomas2023Sci,Zhao2023ApJ}.

Given the extremely high brightness temperature of
FRBs ($\sim 10^{35}~\rm K$), it is expected that
their radiation should be coherent
\citep{Yang2018ApJ,Wang2022ApJ}. The most widely
accepted central engines for FRBs are magnetars
\citep{Metzger2017ApJ,Chime2022Natur.607..256C,Beniamini2023MNRAS}.
Remarkably, the association between FRB 20200428
and an X-ray burst from a galactic magnetar SGR
J1935+2154 directly suggests that at least a
fraction of FRBs originate from magnetized neutron
stars \citep{CHIME2020Natur,Bochenek2020Natur}.
\cite{Zhang2020Natur} divided the FRB models
into two generic categories, which involve
magnetospheres of compact objects (pulsar-like models;
\citealt{Geng2015ApJ,Dai2016ApJ,Kumar2017MNRAS,
Yang2018ApJ,Wadiasingh2019ApJ,Geng2021Innov,Li2021ApJ,
Zhang2022ApJ,Wang2022ApJ,Liu2023ApJ})
or relativistic shocks from the central engines
(GRB-like models; \citealt{Long2018ApJ,Metzger2019MNRAS,
Beloborodov2020ApJ,Margalit2020ApJl}).
However, both kinds of models more or less
encounter some challenges in explaining the complicated
phenomena associated with FRBs \citep{Katz2018MNRAS.481.2946K,Lu2018MNRAS,
Li2021Natur,Nimmo2021NatAs}. Consequently,
despite substantial advancements in observations,
the radiation and triggering mechanisms of FRBs are still elusive
(see \citealt{Zhang2020Natur,Xiao2021SCPMA,Lyubarsky2021Univ,
Zhang2023RvMP}, for reviews).

Extensive efforts have been made to explore the nature of repeating FRBs
(see \citealt{Platts2019PhR} for a review), with
a particular emphasis on sources exhibiting
periodic activities \citep{CHIME2020Natur,
Rajwade2020MNRAS,Geng2021Innov,Li2021ApJ,
Kurban2022ApJ}. FRB models commonly
resort to magnetized neutron stars, which naturally
have an idea energy reservoir (e.g., either rotational
or magnetic energy) and a highly
magnetized plasma environment. Generally, the
periodic activities of some repeating FRBs can be explained
in the framework of two scenarios, e.g., the isolated neutron
star models (spin or free precession;
\citealt{Levin2020ApJ,Beniamini2020MNRAS,
Sob'yanin2020MNRAS,Zanazzi2020ApJ,Xu2021ApJ})
and interacting neutron star models (orbital
modulation; \citealt{Dai2016ApJ,Dai2020ApJ,
Gu2020MNRAS,Decoene2021A&A,Wada2021ApJ,
Kurban2022ApJ,Nurmamat2024EPJC}).

\cite{Geng2015ApJ} suggested that the collision between a neutron
star and an asteroid can produce an FRB event, which could also
potentially explain the association between FRBs and X-ray bursts.
\cite{Dai2016ApJ} argued that when a neutron
star crosses an asteroid belt, repeating FRBs would
be produced by the multiple collision events.
Long-term periodical activities such as the 160-day periodicity
in FRB 20121102A and the 16.35-day periodicity in FRB 20180916B
can be explained as due to the orbital motion of the neutron star
and the asteroid belt. The acceleraton of electrons and
the radiation process are studied in detail \citep{Dai2016ApJ}.
Recently, \cite{Smallwood2019MNRAS} performed
direct $N$-body simulations on the interaction between a
debris disk around a central star and an intruding neutron star.
It is found that most asteroids in the debris disk would be ejected
from the system rather than collide with the neutron star. To
explain the observed FRB rate, an asteroid belt that
is several orders of magnitude denser than
the Kuiper belt would be required \citep{Smallwood2019MNRAS}.

In this study, we investigate the interaction between
a neutron star and an asteroid belt in detail. Long term
dynamical evolution of the debris belt is tracked through
numerical simulations. Various orbital configurations are considered
for the neutron star and the asteroids. The structure
of our paper is organized as follows. Section
\ref{sec2} describes the initial setup of the model.
A brief introduction on the stability criteria is also included.
Numerical results under various orbital conditions are presented
in Section \ref{Numerical result}. Finally, Section \ref{sec:summary}
presents our conclusions and some brief discussion.

\section{Initial setup}
\label{sec2}

The collision between a neutron star and an asteroid/comet could
excite an FRB. Multiple collisions of a series of small objects
with the neutron star would naturally give birth to repeating
bursts \citep{Dai2016ApJ}. During the collision, the induced
electric field will accelerate electrons to ultra-relativistic
speeds, leading to coherent radiation at radio wavelengths and
produce an FRB \citep{Dai2016ApJ}.
\cite{Smallwood2019MNRAS} considered equal-mass binary
neutron star systems, where the asteroid belt is bound to one
binary component. They conducted detailed studies on their
dynamical evolution and collision rates over time. Since a massive
star with a stronger gravity may be able to remain the asteroid
belt bound to it for a much longer time than a 1 -- 2 $M_\odot$
star, it is necessary to explore the cases that the host of the
asteroid belt is a massive star. This may potentially extend the
active timescale of the FRB source and generate periodic FRBs.
Here, we will consider the complicated mechanical interactions in
a multi-body system which constitutes a central massive star, a
belt of asteroids (bound to the massive star), and a
neutron star. The neutron star is assumed to be a companion of the
massive star so that it will traverse the asteroid belt when it
moves in its orbit around the massive star. In our study, the
masses of the neutron star and the central star are fixed as
$m_{\rm NS}=1.4~M_{\odot}$ and $m_{\rm CS}=20~M_{\odot}$,
respectively. The dynamical evolution of the system is then
tracked by means of numerical simulations.

In the collision model, the ultimate energy source is the gravitational
potential energy of the asteroid. As a result, a small object of
$\sim 10^{17}\rm g$ would be sufficient to energize an FRB
\citep{Geng2015ApJ}. This mass is extremely small as compared with
either the neutron star or the massive star. So, in our simulations,
we can safely ignore the gravitational forces from other asteroids
when we analyze the force exerted on a particular asteroid. In other
words, the asteroids are treated as massless particles in our simulations.
It simplifies the complicated system as a restricted
three-body system which includes two main objects and a large number of
test particles.

\subsection{Stability criteria}
\label{stability criteria}

The restricted three-body problem has been extensively studied in
the literature. To get a preliminary impression on the stability of
the asteroid, here we first simplify our system as a triple system
which includes a massive star, an asteroid and a neutron star.
The massive star is located at the center, while both the
asteroid and the neutron star orbit around the massive star with
a semi-major axis of $a_{\rm ast}$ and $a_{\rm NS}$, respectively.
Here we have $a_{\rm NS}>a_{\rm ast}$. The long-term stability of
the asteroid is determined by the system's initial conditions and
can be analyzed based on some simple equations
\citep{Eggleton1995ApJ,Mardling2001MNRAS,He2018mnras}. Following
\cite{Petrovich2015apj}, we define an orbit separation parameter as
\begin{eqnarray} \label{func1}
r_\mathrm{ap}\equiv\frac{a_\mathrm{NS}(1-e_\mathrm{NS})}
{a_\mathrm{ast}(1+e_\mathrm{ast})},
\end{eqnarray}
where $e_{\rm ast}$ and $e_{\rm NS}$ refer to the eccentricities
of the asteroid and neutron star, respectively. A sufficiently
hierarchical triple system ($r_{\rm ap}\gg1$) will be stable on
long timescales and the exchange of angular momentum between the
asteroid and the neutron star occurs without an exchange of energy
\citep{He2018mnras}.

To be more specific, there exists a critical value ($Y$) for the
separation parameter \citep{Mardling2001MNRAS,He2018mnras}, i.e.
\begin{eqnarray} \label{func2}
    Y\equiv2.8\frac{1}{1+e_{\mathrm{ast}}}\left[(1+\mu)
    \frac{1+e_{\mathrm{NS}}}{(1-e_{\mathrm{NS}})^{1/2}}\right]
    ^{2/5}(1-0.3i_{\rm m}/180^\circ),
\end{eqnarray}
where $\mu=\frac{m_{\rm NS}}{m_{\rm CS}}$
represents the mass ratio between the neutron star
and the central massive star, and $i_{\rm m}$ denotes the mutual
inclination between the neutron star's and asteroid's orbits.
If $r_{\rm ap}>Y$, the system will be stable. On the contrary,
when $r_{\rm ap}<Y$, the dynamics of the asteroid becomes
unstable, which means it would either collide with the
massive star/neutron star, or be finally ejected from the system.
From Equation~(\ref{func2}), we see that a smaller $\mu$ (i.e. a
larger $m_{\rm CS}$) and a larger $i_{\rm m}$ will generally make
the system more stable.
However, when the mutual inclination
$i_{\rm m} \gtrsim 39^{\circ}$, the asteroid will be affected by
the Kozai--Lidov (KL) mechanism \citep{kozai1962AJ,Lidov1962P&SS},
leading to an exchange between its eccentricity and orbital inclination.
The maximum eccentricity that an asteroid, initially in a circular
orbit, can achieve is \citep{smallwood2023MNRAS}
\begin{eqnarray} \label{func3}
    e_{\text{ast, max}} = \sqrt{1 - \frac{5}{3} \cos^2 i_{m,0}} ,
\end{eqnarray}
where $i_{\rm m,0}$ denotes the initial mutual inclination. If the
asteroid's orbit is initially inclined at $90^{\circ}$, the KL
effect can excite its eccentricity to be very close to 1. The KL
oscillations may cause the asteroid's periastron to approach the
central massive star, leading to a stellar collision.
Alternatively, they can increase the possibility of a close
encounter with the neutron star due to the increased apastron,
making the asteroid being scattered out of the system.
\citep{Petrovich2015apj}.

\begin{figure}[ht!]
\gridline{\fig{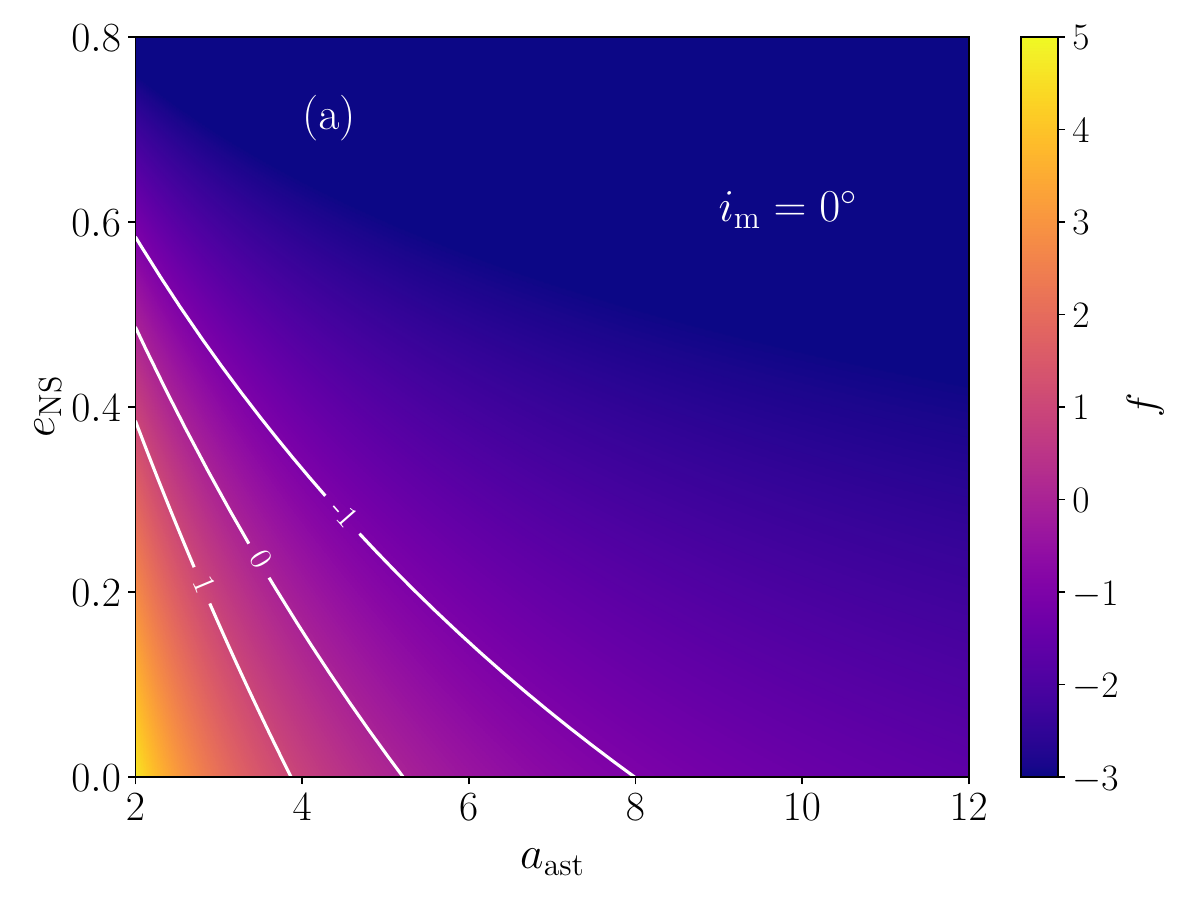}{0.45\textwidth}{}
          \fig{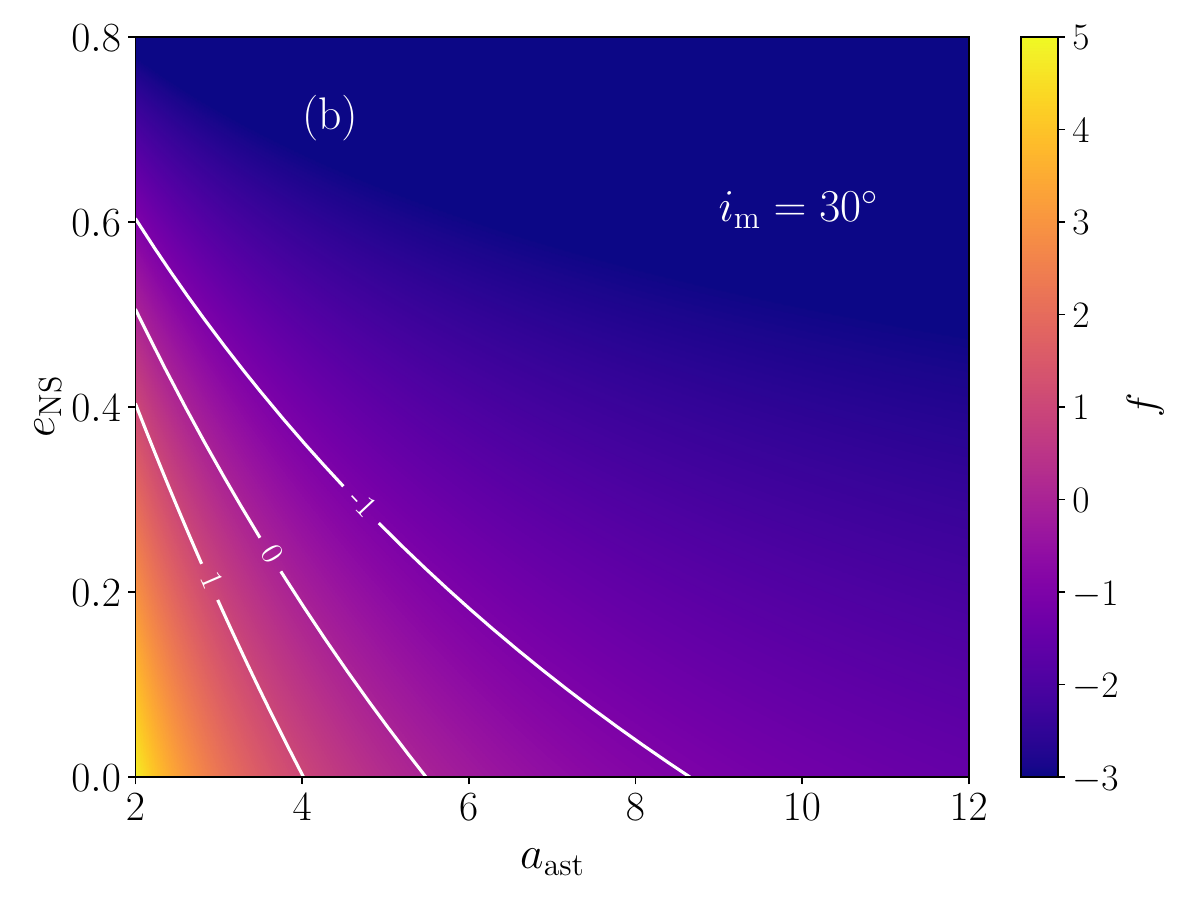}{0.45\textwidth}{}
    }
\caption{Dependence of the stability criterion function $f$
on $a_{\rm ast}$, $e_{\rm NS}$ and $i_{\rm m}$. Panel (a)
illustrates the dependence of $f$ on the semi-major
axis of the asteroid orbit ($a_{\rm ast}$) and the eccentricity
of the neutron star orbit ($e_{\rm NS}$) for a mutual inclination
of $i_{\rm m}=0^{\circ}$. As a comparison, Panel (b) shows the
case of $i_{\rm m}=30^{\circ}$. When $f>0$, the triple
system is stable. The solid curves correspond to isolines of
$f = -1, 0, 1$, respectively.
\label{fig1}}
\end{figure}

We can combine Equation~(\ref{func1}) and Equation~(\ref{func2})
to define a stability criterion function $f$ as
$f \equiv r_{\text{ap}} - Y$. When $f > 0$, the triple system is
dynamically stable, otherwise it is unstable.
Figure~\ref{fig1} illustrates
the dependence of $f$ on $a_{\rm ast}$, $e_{\rm NS}$ and $i_{\rm m}$.
Here we have fixed $e_{\rm ast} = 0$. The semi-major axis of the
neutron star is taken as $a_{\rm NS} = 15$ au, which corresponds
to an orbital period of $P_{\rm orb} = 12.6$ yr. We see that a smaller
$a_{\rm ast}$ and $e_{\rm NS}$ is conducive the stability of the
asteroid.
Compared to the zero inclination cases, the systems
with $i_{\rm m} = 30^{\circ}$ allow for long-term stable asteroids
with slightly larger semi-major axes and eccentricities.
Again, asteroids with an initial mutual inclination in
the KL oscillation region (i.e. $39^{\circ} \lesssim i_{\rm m}
\lesssim 141^{\circ}$) will be subject to KL instabilities, which
makes it difficult to judge the stability of the system. In other
words, the criteria are ineffective on predicting the long-term
stability of highly inclined systems \citep{Petrovich2015apj}.
The stability analysis can guide us in evaluating the various
parameters involved in our modeling.


\subsection{Configurations and initial setup}
\label{intial parameters}

We have performed numerical simulations on the dynamical evolution
of the asteroid belt in a massive star-neutron star binary.
Various configurations are considered for the system. According to
the relative position of the neutron star and the asteroid belt,
our simulations can be grouped into two scenarios. In the first
scenario (Scenario I), the orbit of the neutron star will cross the
asteroid belt, while in the second scenario (Scenario II), the orbit
of the neutron star is encompassed by the asteroid belt. In all the
cases, the mass of the central massive star is taken
as $m_{\rm CS}=20~M_{\odot}$. The neutron star has a mass
of $m_{\rm NS}=1.4~M_{\odot}$, and it moves in an orbit with a
semi-major axis of $a_\mathrm{NS} = 15$ au. Four different
values are taken for the eccentricity of the neutron star orbit, i.e.
$e_{\rm NS} =$ 0, 0.3, 0.5, 0.7. 20000 asteroids are assumed to orbit
around the massive star in a belt. Their initial eccentricities
($e_{\rm ast}$) are randomly evaluated in a range of 0 -- 0.1.
They are not strictly coplanar, instead, their initial inclination
angles ($i_{\rm ast}$) are randomly distributed between 0 and 0.1 rad.
Other angular parameters of the asteroids, such as the argument of
pericenter ($\omega_{\rm ast}$), the longitude of the ascending node
($\Omega_{\rm ast}$) and the initial true anomaly ($M_{\rm ast}$),
are randomly assigned in the phase range of 0 -- $2\pi$.

\begin{figure}[ht!]
\gridline{\fig{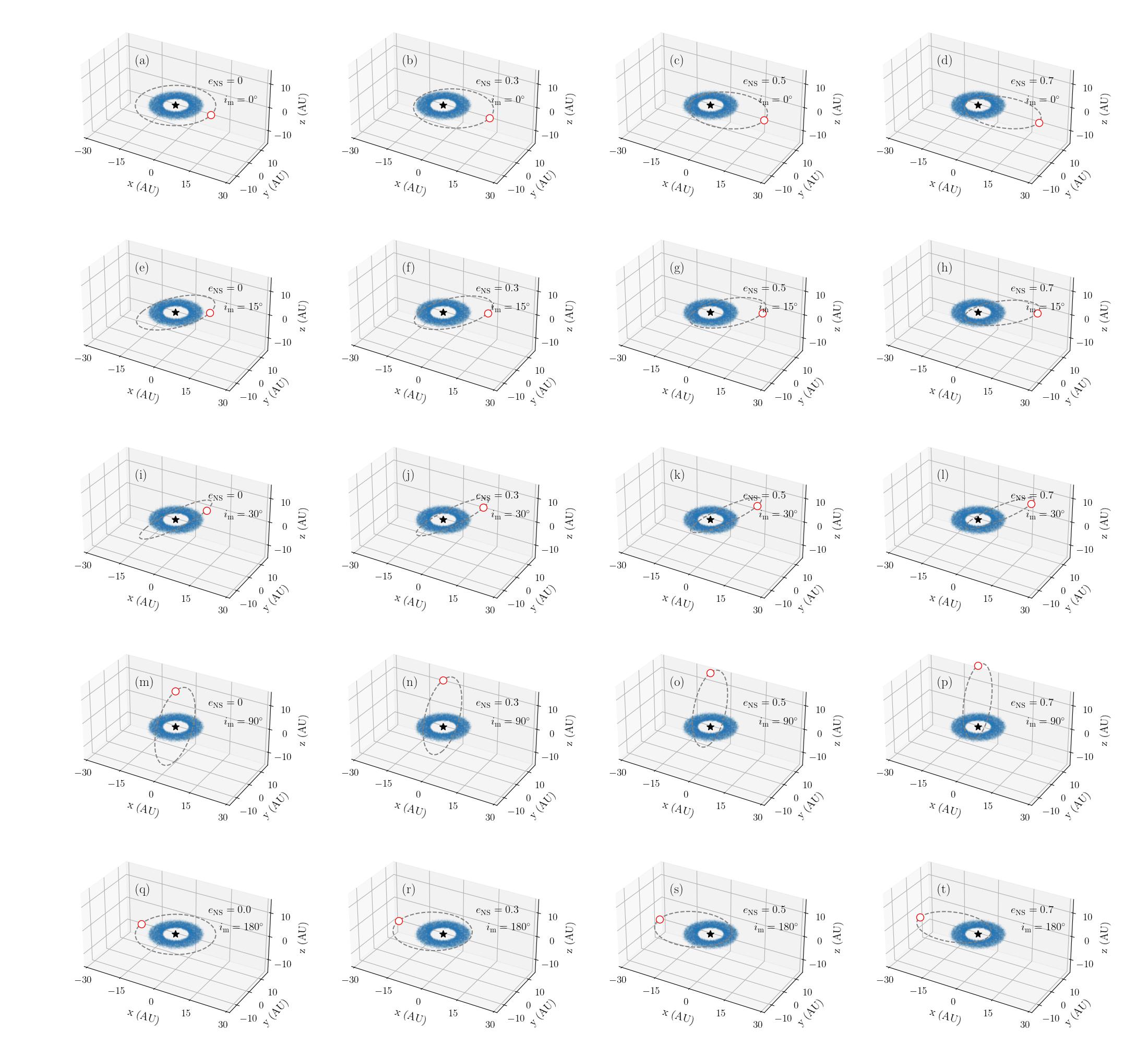}{1\textwidth}{}}
\caption{A schematic illustration of the initial conditions of
the 20 cases in Scenario \Rmnum{1}. 20000 small objects
are included in the asteroid belt, with the semi-major axis ranging
in 5 au -- 10 au. The semi-major axis of the neutron star orbit
is $a_\mathrm{NS} = 15$ au and the eccentricity is taken as
$e_{\rm NS} =$ 0, 0.3, 0.5 or 0.7. Five angles are assumed
for the mutual inclination between the main plane of the asteroid
belt and the neutron star orbit, i.e. $i_{\rm m} = 0^{\circ}$,
$15^{\circ}$, $30^{\circ}$, $90^{\circ}$, or $180^{\circ}$. The
dotted line represents the trajectory of the neutron star. The
central massive star, neutron star and asteroids are represented
by the star symbol, the hollow circle and the dots, respectively
(size not to scale).
\label{fig2}}
\end{figure}

In Scenario I, the initial semi-major axes ($a_{\rm ast}$) of the 20000
asteroids are taken randomly in a range of 5 au -- 10 au. They are
generally smaller than $a_\mathrm{NS}$ so that the neutron star will
essentially travel through the asteroid belt and interact strongly with
them when it orbits around the massive star. Different mutual
inclination angles ($i_{\rm m}$) between the main plane of the asteroid
belt and the neutron star orbit are assumed in our calculations,
i.e. $i_{\rm m} = 0^{\circ}$, $15^{\circ}$, $30^{\circ}$, $90^{\circ}$,
and $180^{\circ}$. Note that a mutual inclination of $180^{\circ}$
means that the neutron star is essentially in a retrograde orbit.
The five inclination angles combined with the four eccentricities
of the neutron star orbit indicate that there are 20 cases in Scenario
\Rmnum{1}, of which the initial conditions are illustrated in
Figure \ref{fig2}.

\begin{figure}[ht!]
\gridline{\fig{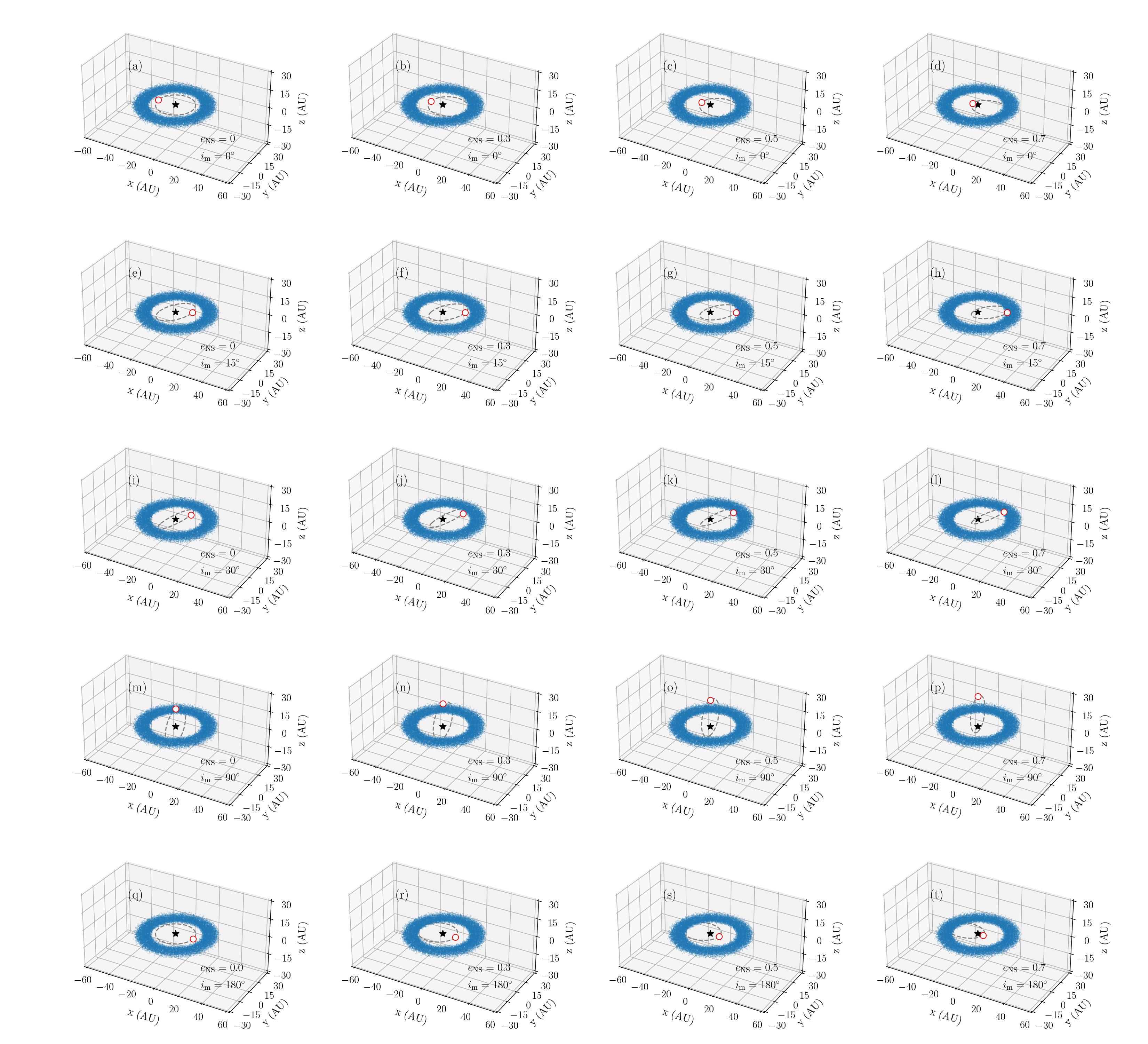}{1\textwidth}{}}
\caption{A schematic illustration of the initial conditions of
the 20 cases in Scenario \Rmnum{2}. 20000 small objects
are included in the asteroid belt, with the semi-major axis
ranging in 20 au -- 30 au. The semi-major axis of the neutron
star orbit is $a_\mathrm{NS} = 15$ au and the eccentricity
is taken as $e_{\rm NS} =$ 0, 0.3, 0.5 or 0.7. Five
angles are assumed for the mutual inclination between the main
plane of the asteroid belt and the neutron star orbit, i.e.
$i_{\rm m} = 0^{\circ}$, $15^{\circ}$, $30^{\circ}$, $90^{\circ}$,
or $180^{\circ}$. Line styles and symbols are the same as those
in Figure \ref{fig2}.
\label{fig3}}
\end{figure}

The setup in Scenario \Rmnum{2} corresponds to a
circumbinary disk orbiting around the binary system. Observations
indicate that misaligned discs around binary system are common
\citep{Brinch2016,Kennedy2019NatAs}. Following
\cite{Chen2020MNRAS}, we set the particles in the circumbinary
disks to move in orbits with respect to the center of mass of the
binary system. The semi-major axes of the asteroids are randomly
distributed between 20 and 30 au, while $a_{\rm NS}$ is still 15
au. As a result, the neutron star orbit is generally encompassed
by the asteroid belt. The eccentricity of the neutron star
orbit is still taken as $e_{\rm NS} = $ 0, 0.3, 0.5 or 0.7, and
the mutual inclination is evaluated as $i_{\rm m} = 0^{\circ}$,
$15^{\circ}$, $30^{\circ}$, $90^{\circ}$, or $180^{\circ}$. So, we
also have a total number of 20 cases for Scenario \Rmnum{2}. The
initial conditions of these configurations are illustrated in
Figure \ref{fig3}.

We employ the IAS15 integrator \citep{Rein2015MNRAS} to perform numerical
simulations of our systems. It is an adaptive method which varies the
step size automatically and is thus quite suitable for simulations that
involve close encounters. Note that the IAS15 integrator is
included in the open-source N-body code \texttt{REBOUND}
\citep{Rein2012A&A}. Our numerical results are presented
in Section \ref{Numerical result} below.

\subsection{Collision and ejection of the asteroids}

When a neutron star travels through an asteroid belt, its collision
rate with the asteroids can be simply estimated as \citep{Dai2016ApJ}
\begin{eqnarray} \label{func4}
\mathcal{R}_{\rm a}=\sigma_{\rm a}v_\ast n_{\rm a},
\end{eqnarray}
where $v_\ast$ is the proper velocity of the neutron star and
$n_{\rm a}$ is the number density of asteroids in the belt. $\sigma_{\rm a}$
is the impact cross-section given by \citep{Smallwood2019MNRAS}
\begin{eqnarray} \label{func5}
\sigma_{\rm a}=\frac{4\pi Gm_{\rm NS}R_{*}}{v_{*}^{2}},
\end{eqnarray}
where, $G$ and $R_\ast$ are the gravitational constant
and neutron star radius, respectively. In our framework,
gravitational interactions between the neutron star and the asteroid
belt significantly alters the shape of the belt, which means the
number density of asteroids and thus the collision rate evolves
rapidly with time. We then need to promptly examine when and where an
asteroid will collide with the neutron star in the simulation.

An asteroid will be tidally disrupted by the neutron star
when it approaches the tidal disruption radius defined by
\citep{Hills1975Natur,Wu2023MNRAS}
\begin{eqnarray} \label{func6}
R_{\rm t,NS} \approx (6m_{\mathrm{NS}}/\pi\rho)^{1/3} \sim 8.1 \times 10^{10}
(m_{\mathrm{NS}}/1.4M_{\odot})^{1/3}\rho_{1}^{-1/3}\mathrm{~cm},
\end{eqnarray}
where the convention $Q_x=Q/10^x$ in units of cgs is adopted. For a
homogeneous iron-nickel asteroid, the density is typically $\sim 8~\rm g~cm^{-3}$.
Note that the strong magnetic field of the neutron star will further help
capture the asteroid, leading to a collision event. Here we
take $R_{\rm t,NS}$ as a conservative criterion for collision.
In our simulations, when the separation
between an asteroid and the neutron star decreases to $R_{\rm t,NS}$,
we consider it to essentially collide with the compact object, which may
effectively lead to an FRB.
Note that some asteroids may even collide with the central massive star due
to the gravitational perturbations from the neutron star. Similarly, we also
take the tidal disruption radius as the criterion for collision with the
central massive star, i.e.
$R_{\rm col,CS} = (6m_{\mathrm{CS}}/\pi\rho)^{1/3} =
(m_{\rm CS}/m_{\rm NS})^{1/3} R_{\rm t,NS}$.

In addition to colliding with the massive star/neutron star, some
asteroids may also be ejected out of the system \citep{Smallwood2019MNRAS}.
In Scenario \Rmnum{1}, an ejection event is identified when the
distance of an asteroid (with respect to the central massive star)
is four times larger than the semi-major axis of the neutron star
(60 au). In Scenario \Rmnum{2} the ejection criterion is
taken as 100 au. We cease to track the asteroid in our simulations
as long as it is identified as being ejected.

\section{Numerical results}
\label{Numerical result}

The evolution of the systems is tracked till the neutron
star moves around the central massive star for 300 times in the
orbit (i.e. after 300 orbital periods) for both Scenarios
\Rmnum{1} and \Rmnum{2}.

\subsection{Scenario \Rmnum{1}: neutron star being on an external trajectory}
\label{sec3.1}

\begin{figure}[ht!]
\gridline{\fig{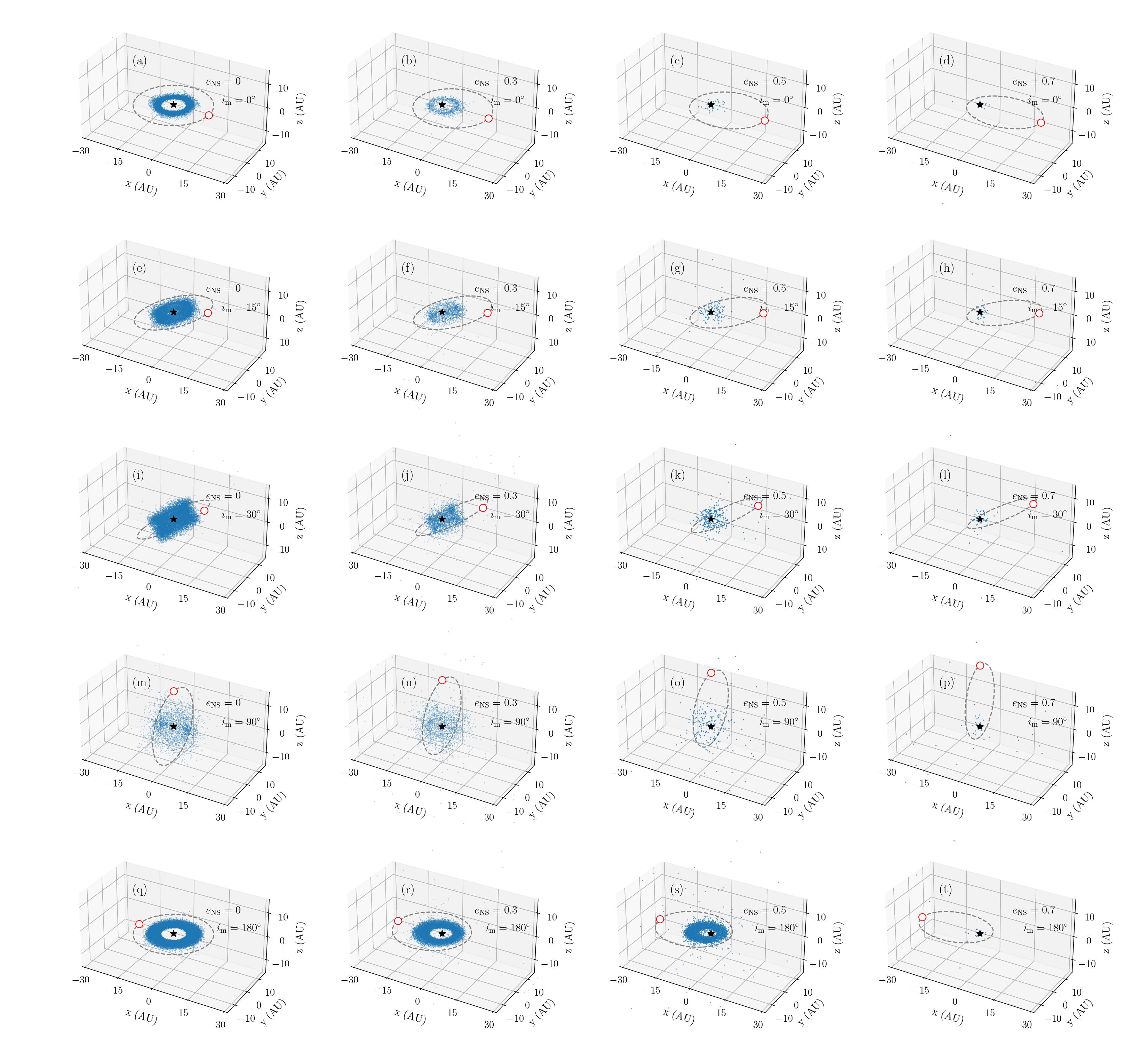}{1\textwidth}{}}
\caption{The spatial distribution of the asteroids after 300
periods of evolution (i.e. at the moment of $t = 300~P_{\rm orb}$),
for the 20 cases of Scenario \Rmnum{1} with the initial
conditions illustrated in Figure \ref{fig2} correspondingly.
Line styles and symbols are the same as in Figure \ref{fig2}.
\label{fig4}}
\end{figure}

Figure \ref{fig4} shows the spatial distribution of the asteroids
at the final stage of our simulation (i.e. at $t$ = 300
$P_{\rm orb}$) for the 20 cases of Scenario \Rmnum{1}
(whose initial distributions are illustrated in Figure \ref{fig2},
correspondingly). We see that the basic structure of the
asteroid belt is better preserved in cases with $i_{\rm m} =
180^{\circ}$ compared to those with $i_{\rm m} = 0^{\circ}$ when
the eccentricity of the neutron star is small (e.g. $e_{\rm NS} =
$ 0 or 0.3). As the eccentricity increases, the asteroid belt is
more seriously depleted after 300 $P_{\rm orb}$. The reason is
that a larger $e_{\rm NS}$ means a smaller periastron for the NS
trajectory so that the NS orbit penetrates deeper into the
asteroid belt, leading to a stronger gravitational interaction
between the neutron star and the asteroids. Especially, in
all cases of $e_{\rm NS}=0.7$, the vast majority of asteroids
either are ejected from the system or collide with the binary
stars, with only a small portion remain bound in the system at $t$
= 300 $P_{\rm orb}$. For low-inclination  systems (e.g.
$i_{\rm m}\leqslant 30^{\circ}$) although a nonzero mutual
inclination adds extra stability to the system, the asteroids are
significantly scattered by the NS and are thus dispersed
throughout the entire 3D space. It significantly reduces the
number density of asteroids along the path of the neutron star
orbit, therefore decreases the collision rate with the
neutron star. As the inclination increases to
$90^{\circ}$, the highly tilted orbits excite strong KL
oscillations \citep{smallwood2023MNRAS}, also making the system
unstable.

\begin{figure}[ht!]
\gridline{\fig{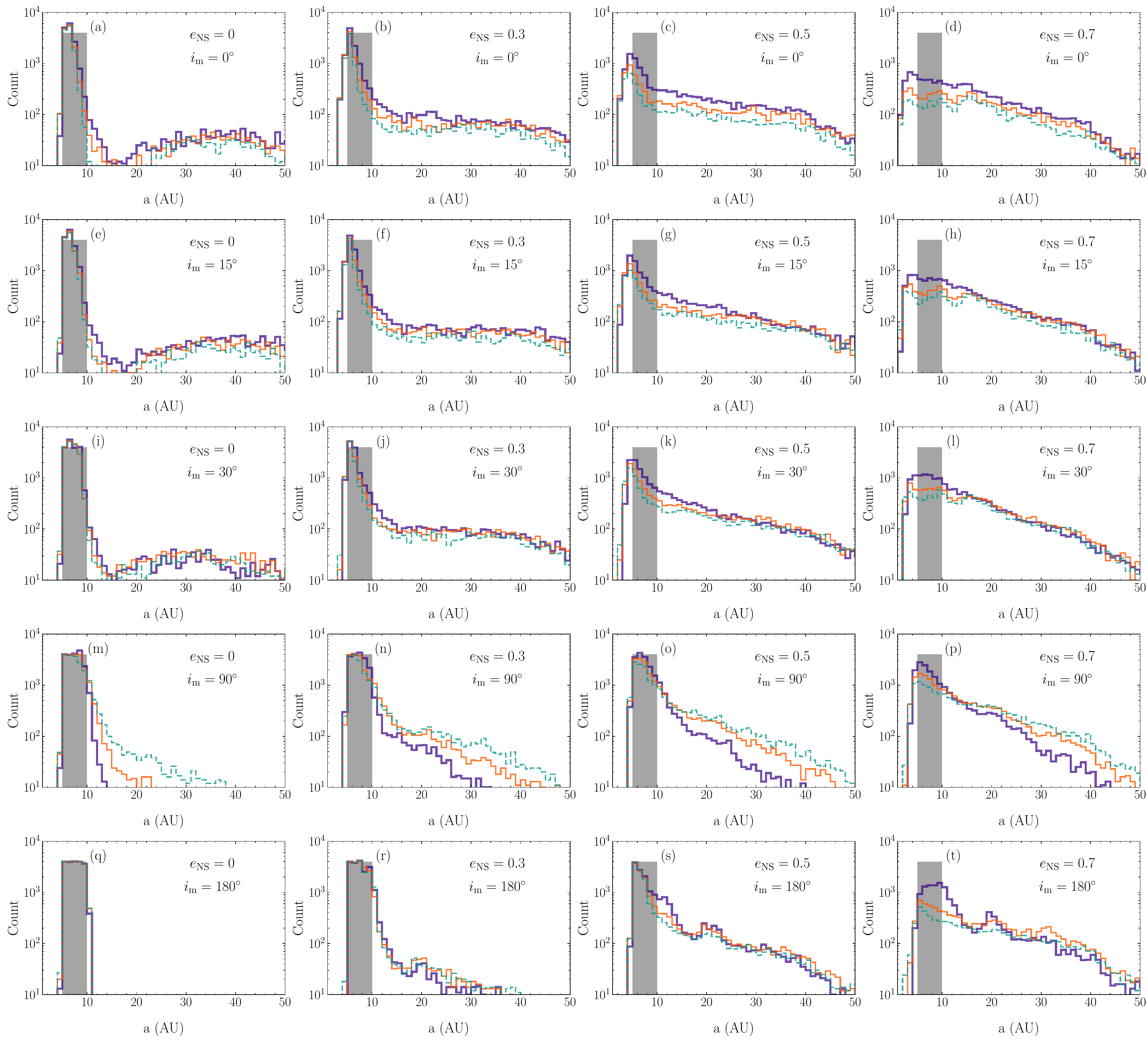}{1\textwidth}{}}
\caption{The evolution of the semi-major axis distribution of the
asteroids for the 20 cases of Scenario \Rmnum{1} (the
initial conditions are illustrated in Figure \ref{fig2} correspondingly).
In each panel, the thick solid line, thin solid line, and dashed line
represent the distribution at $t = 5~P_{\rm orb}$, $10~P_{\rm orb}$,
and $15~P_{\rm orb}$, respectively. The gray shaded rectangle indicates
the initial semi-major axis distribution of the asteroids (i.e. at
$t = 0~P_{\rm orb}$).
\label{fig5}}
\end{figure}

\begin{figure}[ht!]
\gridline{\fig{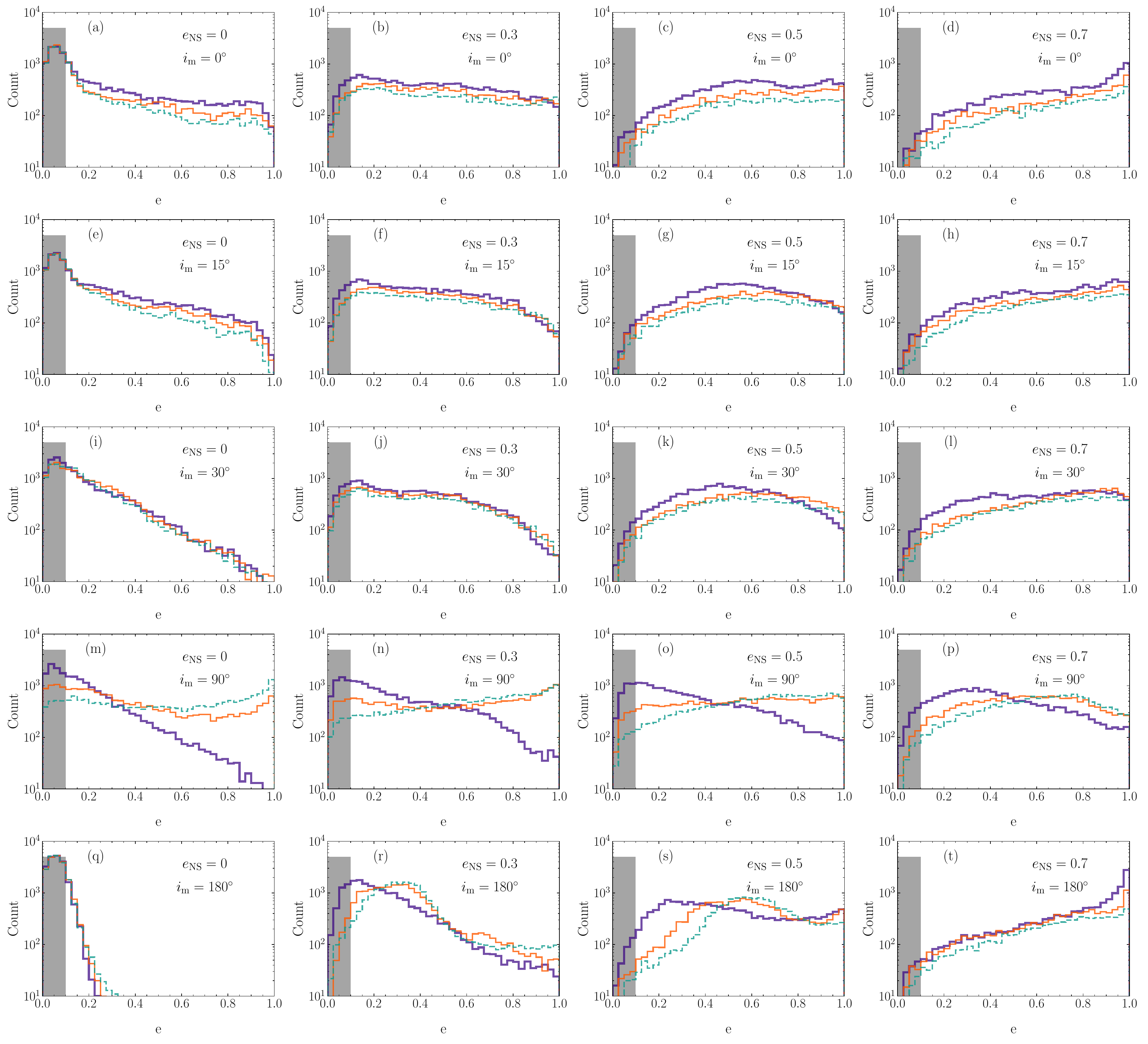}{1\textwidth}{}}
\caption{The evolution of the eccentricity distribution of the
asteroids for the 20 cases of Scenario \Rmnum{1} (the
initial conditions are illustrated in Figure \ref{fig2} correspondingly).
In each panel, the thick solid line, thin solid line, and dashed line
represent the distribution at $t = 5~P_{\rm orb}$, $10~P_{\rm orb}$,
and $15~P_{\rm orb}$, respectively. The gray shaded rectangle indicates
the initial eccentricity distribution of the asteroids (i.e. at
$t = 0~P_{\rm orb}$).
\label{fig6}}
\end{figure}

The evolution of the semi-major axis distribution of the asteroids
is shown in Figure \ref{fig5} for the 20 cases of
Scenario \Rmnum{1}. Correspondingly, Figure \ref{fig6} illustrates
the evolution of the eccentricity distribution. The distributions
at three moments, i.e. $t$ = 5 $P_{\rm orb}$, 10 $P_{\rm orb}$,
and 15 $P_{\rm orb}$ are plot, offering a clear representation of
their temporal evolution.
The semi-major axes and
eccentricities of the asteroids are measured relative to the
central massive star. We mainly focus on the evolution of
asteroids bound to the central massive star, which have negative
specific energy ($e_{\rm ast} < 1$). The asteroids with positive
specific energy ($e_{\rm ast} > 1$) are not included in Figures
\ref{fig5} and \ref{fig6}. Figure \ref{fig5} further confirms
that an increase in $e_{\rm NS}$ will lead more asteroids to be
ejected out of the system, which is consistent with the picture
shown in Figure \ref{fig4}. Concurrently, the remaining asteroids
exhibit a broader spatial distribution and a preference for higher
eccentricities. In other words, reducing the orbital eccentricity
of the neutron star can alleviate the destroy to the asteroid belt
and prolong the lifespan of the system. As long as $i_{\rm m}$ is
not too large (e.g. $i_{\rm m}=15^{\circ}$ and $30^{\circ}$), the
mutual inclination has a very limited impact on the evolution
pattern of the semi-major axes and eccentricities. Note that a
large fraction of the asteroids are actually scattered into the 3D
space (see Figure \ref{fig4}), which is not reflected by the
distributions of the two parameters.
When the neutron star is
in a retrograde circular orbit, it has a relatively minor
influence on the asteroid belt. Only a small fraction of the
particles are affected, whose eccentricities increase slightly. 
In the $i_{\rm m}=90^{\circ}$ cases, we can clearly see
the KL oscillations from the evolution of the asteroids'
eccentricity distribution. Theoretically, the KL oscillation
period of an asteroid in an eccentricity binary can be estimated
as \citep{smallwood2023MNRAS}
\begin{eqnarray} \label{func7}
\frac{\tau_{\rm KL}}{P_{\rm orb}} \approx \frac{M_{\rm CS} + M_{\rm NS}}{M_{\rm NS}}
\frac{P_{\rm orb}}{P_{\rm ast}} \left(1 - e_{\rm NS}^2\right)^{3/2},
\end{eqnarray}
where $P_{\rm ast}=2\pi/ \sqrt{GM_{\rm CS}/a_{\rm ast}^{3}}$ is
the period of the asteroid orbiting around the central massive
star. Equation (\ref{func7}) indicates that an asteroid with a
shorter period exhibits a longer KL oscillation period. In our
cases, the upper limit of the KL oscillation period should be
$\tau_{\rm KL, max} \approx 76~P_{\rm orb}$. In Figure \ref{fig6}
(m) -- (p), when $t \geqslant 10~P_{\rm orb}$, the eccentricities
of asteroids continuously increase over time, some are even
excited by the KL mechanism to approach 1 (see Equation
(\ref{func3})). As a result, the significantly increased
eccentricity will lead the periastron of the asteroids to be very
close to the central massive star, increasing the possibility of
collision. Especially, in the case with $i_{\rm m}=90^{\circ}$ and
$e_{\rm NS} = 0$, the KL oscillation causes $\sim$ 40 percent of
the asteroids to collide with the central star in $t < 300~P_{\rm
orb}$, which is the highest portion among the 20 cases in Scenario
\Rmnum{1}.

\begin{figure}[ht!]
\gridline{\fig{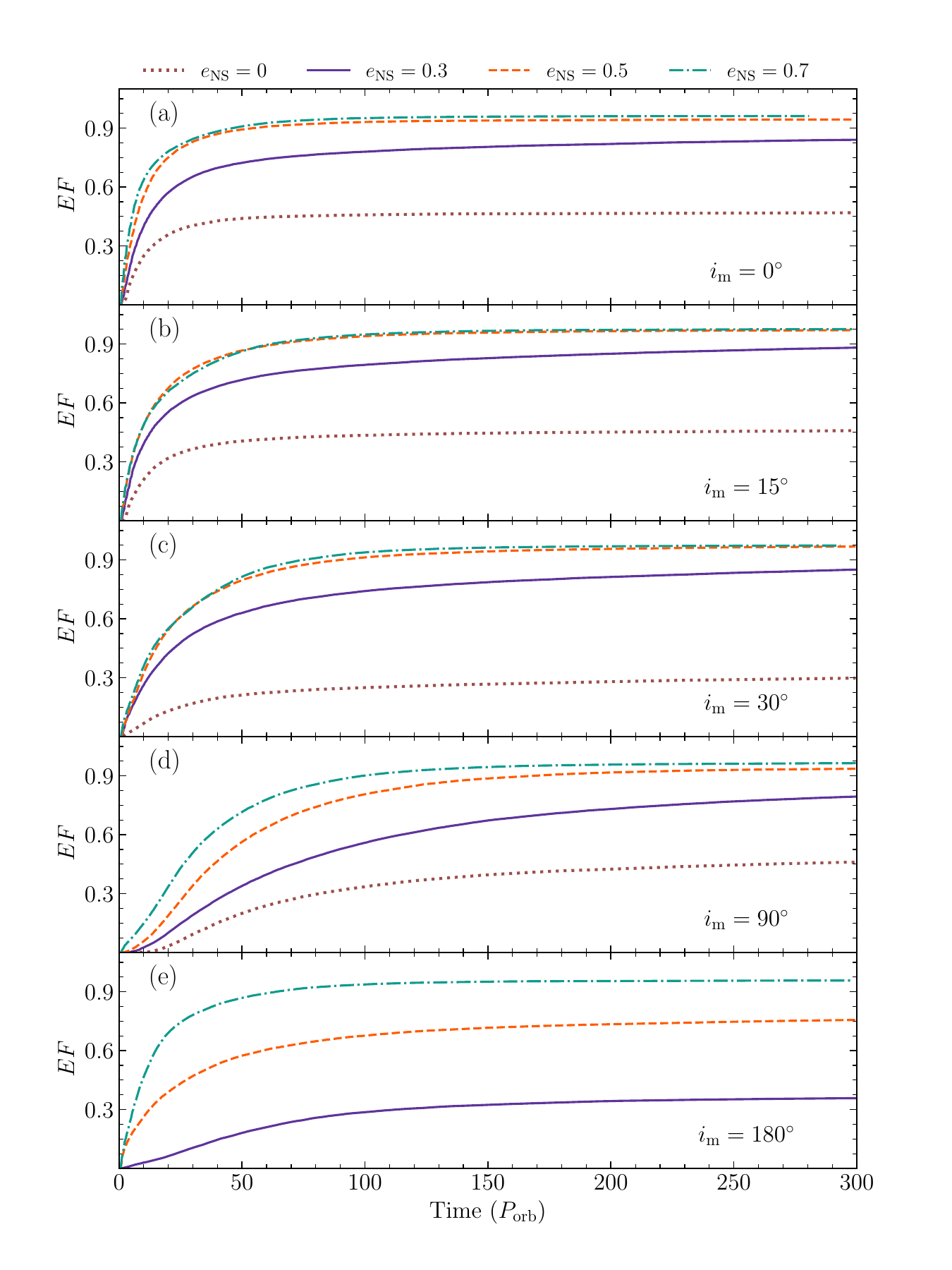}{0.8\textwidth}{}}
\caption{The cumulative ejection fraction ($EF$) of the asteroids as
a function of time for the 19 cases in Scenario \Rmnum{1}.
Note that in the case with $i_{\rm m}=180^{\circ}$ and
$e_{\rm NS}=0$, none of the particles is ejected from the system.
Time is measured in units of the neutron star's orbital period
($P_{\rm orb}$). The five panels from top to bottom correspond to
the mutual inclination of $i_{\rm m}=0^{\circ}$, $15^{\circ}$,
$30^{\circ}$, $90^{\circ}$, and $180^{\circ}$, respectively.
The dotted line, solid line, dashed line, and dash-dotted line
in each panel represent $e_{\rm NS}$ = 0, 0.3, 0.5, and
0.7, respectively.
\label{fig7}}
\end{figure}

The cumulative ejection fraction ($EF$) of the asteroids is plot
versus time in Figure \ref{fig7}.
In almost all cases, the
evolution of $EF$ is levelled off at $t=300~P_{\rm orb}$,
indicating that the system has entered a relatively stable stage.
An increase in the mutual inclination slightly reduces the early
growth rate of $EF$. On the other hand, a larger $e_{\rm NS}$
significantly increases $EF$. In the cases with a non-zero $e_{\rm
NS}$ and $i_{\rm m} \leqslant 90^{\circ}$, we have $EF \sim $ 0.75
-- 0.9 at the final stage. The neutron star with a low
eccentricity in a retrograde orbit scatters much fewer asteroids
out of the system (see also Figure \ref{fig4} -- \ref{fig6}) as
compared with other cases. In our calculations, no asteroids are
recorded to collide with the binary or are ejected from the system
in the $i_{\rm m}=180^\circ$, $e_{\rm NS}=0$ case.

\begin{figure}[ht!]
\gridline{\fig{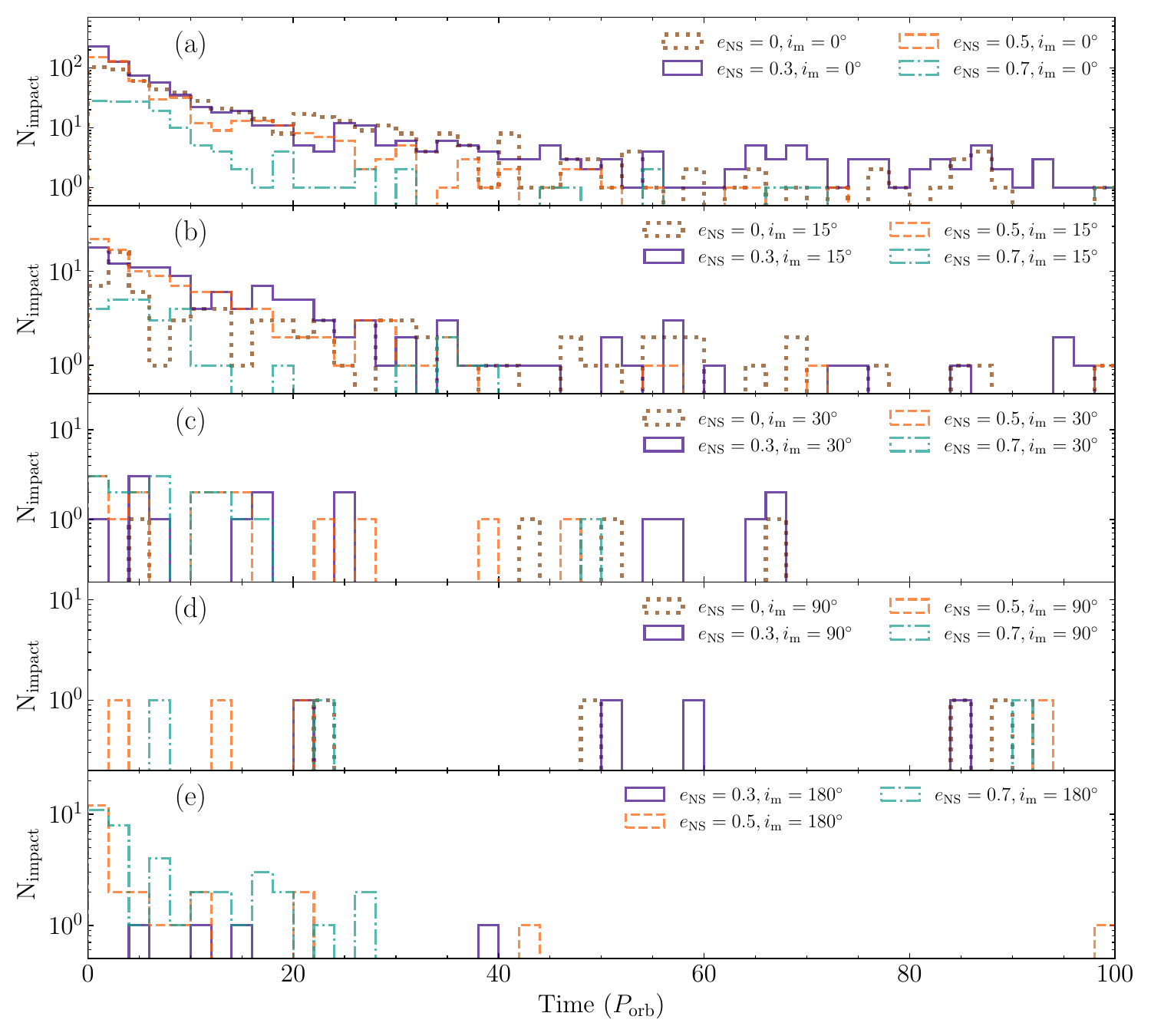}{1\textwidth}{}} \caption{The impact
number versus time for the 19 cases of Scenario
\Rmnum{1}. Note that in the case with $i_{\rm
m}=180^{\circ}$ and $e_{\rm NS}=0$, none of the particles collide
with the neutron star. Time is measured in units of the neutron
star's orbital period $P_{\rm orb}$. Panels (a) -- (e) correspond
to the mutual inclination of $i_{\rm m}=0^{\circ}$, $15^{\circ}$,
$30^{\circ}$, $90^{\circ}$, and $180^{\circ}$, respectively. The
dotted line, solid line, dashed line, and dash-dotted line in each
panel represent $e_{\rm NS}$ = 0, 0.3, 0.5, and 0.7,
respectively. \label{fig8}}
\end{figure}

Figure \ref{fig8} shows the temporal evolution of the impact
rate for the 20 cases in Scenario \Rmnum{1}.
The highest impact rate is observed in the case with
$i_{\rm m}=0^\circ$ and $e_{\rm NS}=0.3$. The impact rate
averaged over first 50 $P_{\rm orb}$ is
$\sim 1.2~\rm year^{-1}$ with a belt number density of
$n_{\rm a} \sim N_{\rm ast}/{V_{\rm ast}} \sim 42~ \rm au^{-3}$.
The average collision rate would be further elevated if one
solely concentrates on an early stage (e.g. $t<20~P_{\rm orb}$).
Increasing the belt number density will correspondingly increase
the collision rate (e.g. \citealt{Smallwood2019MNRAS}).
The increase of the neutron star eccentricity ($e_{\rm NS}$)
not only diminishes the impact rate, but also effectively
disperses the asteroids from the system (see Figure \ref{fig7}).
A nonzero mutual inclination can suppress incidence of
collision events, even at a relatively low inclination
(e.g. $i_{\rm m}=15^{\circ}$).
This is attributed to the low number density along the orbital path
of the neutron star when its orbit is not coplanar
with the main plane of the asteroid belt. Note that in the coplanar
cases (i.e. $i_{\rm m}=0^{\circ}$ and $180^{\circ}$), the
neutron star with $i_{\rm m}=0^{\circ}$ is more adept at inducing
collision events compared to the one with $i_{\rm m}=180^{\circ}$.
For the cases with considerable collision events
(i.e. $i_{\rm m}=0^{\circ}$ and $15^{\circ}$ cases), the impacts
primarily occur in the first 20 $P_{\rm orb}$. So, in this
framework, the lifespan of the FRB source is primarily determined
by the neutron star's orbital period. A natural inference is that
the activity of such FRB sources will gradually decrease over time,
owing to the diminished number density within the belt due to the
neutron star's multiple passages.

\begin{figure}[ht!]
\gridline{\fig{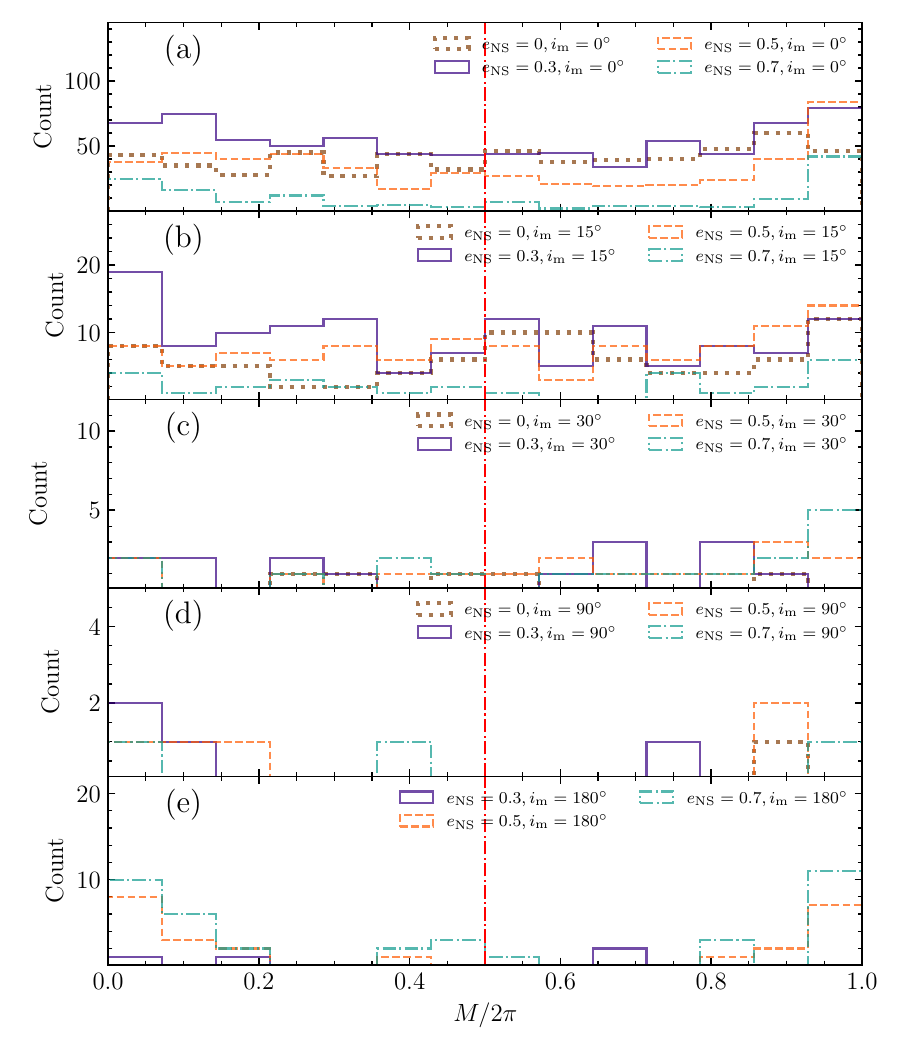}{0.8\textwidth}{}} \caption{The phase
distribution of the collision events when they are folded
according to the neutron star orbital period. Here the Y-axis is
the number of collision events and the X-axis is the mean anomaly
$M$ of the neutron star. The 19 cases of Scenario
\Rmnum{1} are shown. Note that in the case with $i_{\rm
m}=180^{\circ}$ and $e_{\rm NS}=0$, none of the particles collide
with the neutron star. Panels (a) -- (e) correspond to the mutual
inclination of $i_{\rm m}=0^{\circ}$, $15^{\circ}$, $30^{\circ}$,
$90^{\circ}$, and $180^{\circ}$, respectively. Note that $M = 0$
and $2\pi$ corresponds to the periastron of the neutron star
orbit, and the vertical dash-dotted line at $M = \pi$ represents
the apastron. The dotted line, solid line, dashed line, and
dash-dotted line in each panel represent $e_{\rm NS}$ =
0, 0.3, 0.5, and 0.7, respectively. \label{fig9}}
\end{figure}

An interesting problem is whether there are any periodicity in
the collision activities. To explore the periodicity, we have
folded the collision events with respect to the neutron star
orbital period. Figure \ref{fig9} plots the counts of collision
events versus the  mean anomaly $M$ of the neutron star for the
20 cases of Scenario \Rmnum{1}. The mean anomaly $M$ is
a linear function of time, which is given by
\begin{eqnarray} \label{func8}
M=M_{\rm 0}+nt,
\end{eqnarray}
where $M_{\rm 0}$ is the initial value of the mean anomaly at
$t=0~P_{\rm orb}$ and $n=2\pi/P_{\rm orb}$ is the mean motion.
$M=0$ indicates that the neutron star is at the periastron. In the
$i_{\rm m}=0^{\circ}$ and $e_{\rm NS} = 0.7$ case, the collision
events mainly happen in a phase range of $M/2\pi$ = 0.9 --- 1,
suggesting a clear periodic signal in the collisions. It indicates
that the collisions predominantly occur shortly before the neutron
star reaches its periastron.  We notice that the distribution of
the collision events is not symmetric at around the periastron. It
implies that the shape of the asteroid belt is significantly
altered as the neutron star passes through it. In the $i_{\rm
m}=0^{\circ}$ and $e_{\rm NS} = 0.5$ case, although there are
still many collision events in the phase range of $M/2\pi$ = 0.9
--- 1, we notice that there are also many events occurring at
other phases. As $e_{\rm NS}$ further decreases (i.e. $e_{\rm
NS}=$ 0.3 or 0, and $i_{\rm m}=0^{\circ}$), the
distribution of the collision events becomes approximately uniform
in the whole phase range. We can use the reduced chi-square
$\chi^{2}_{\rm red}$ to quantify the extent to which a
distribution deviates from a uniform distribution.
The reduced $\chi^{2}_{\rm red}$ of the above four cases are 11.8,
9.6, 3.5, and 1.1, respectively, which decreases as $e_{\rm NS}$
decreases. Again, we see that the $e_{\rm NS}=0.7$ case exhibits
the highest $\chi^{2}_{\rm red}$, further demonstrating that a
periodic pattern exists in the collision activities. In the cases
with $i_{\rm m} = 15^{\circ}, 30^{\circ}, 90^{\circ}$, the phase
distribution is generally consistent with a uniform distribution,
suggesting the lack of any periodicity. In the cases of $i_{\rm
m}=180^{\circ}$, the collision events also mainly happen near the
phase of  $M/2\pi$ = 1 when $e_{\rm NS}=0.5$ or $e_{\rm NS}=0.7$,
suggesting that the collisions might exhibit periodic behavior.
However, the event rate is generally very low in our simulations,
making it challenging to obtain a quantitative demonstration
through the $\chi^{2}$ test.

\begin{figure}[ht!]
\gridline{\fig{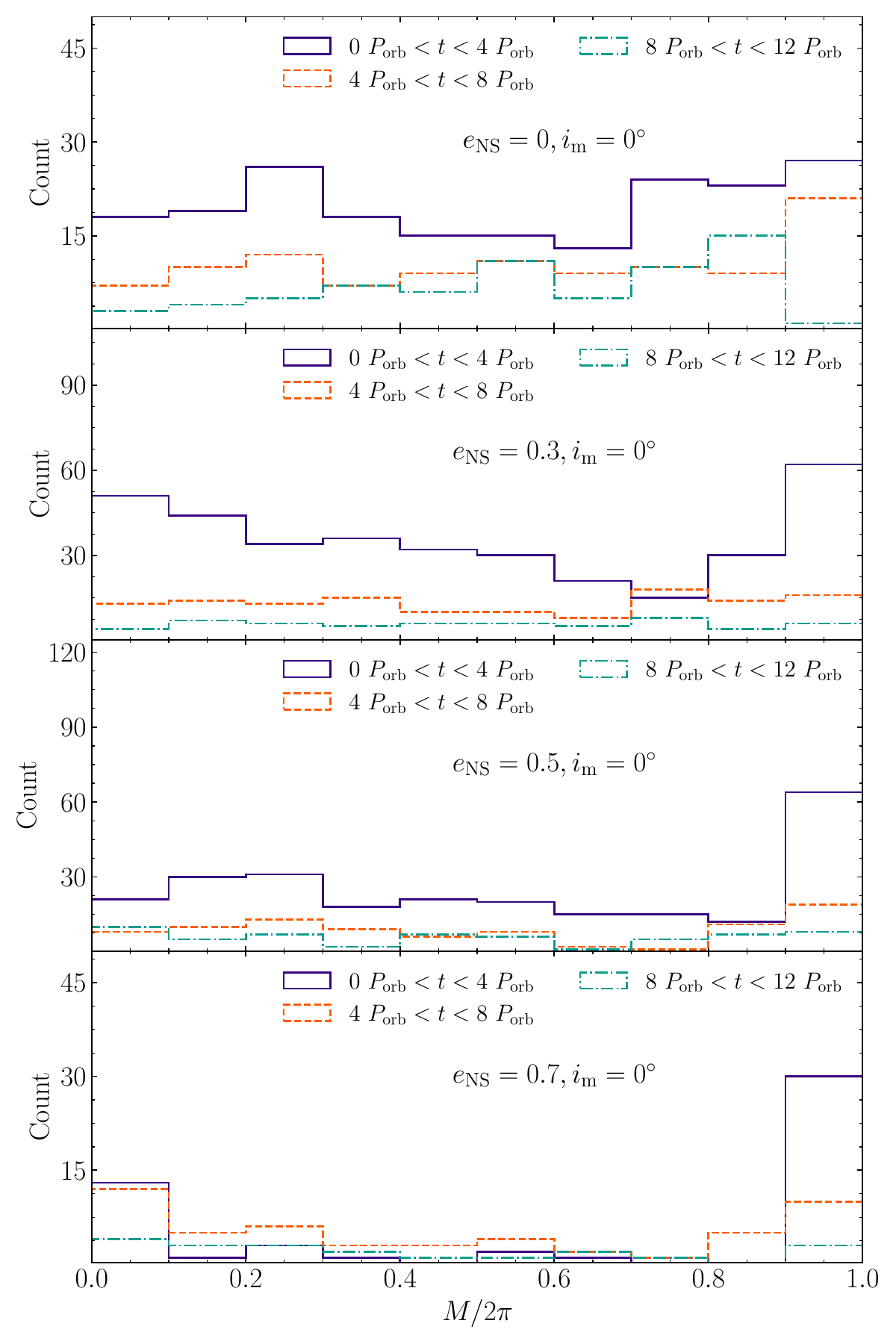}{0.47\textwidth}{}
         \fig{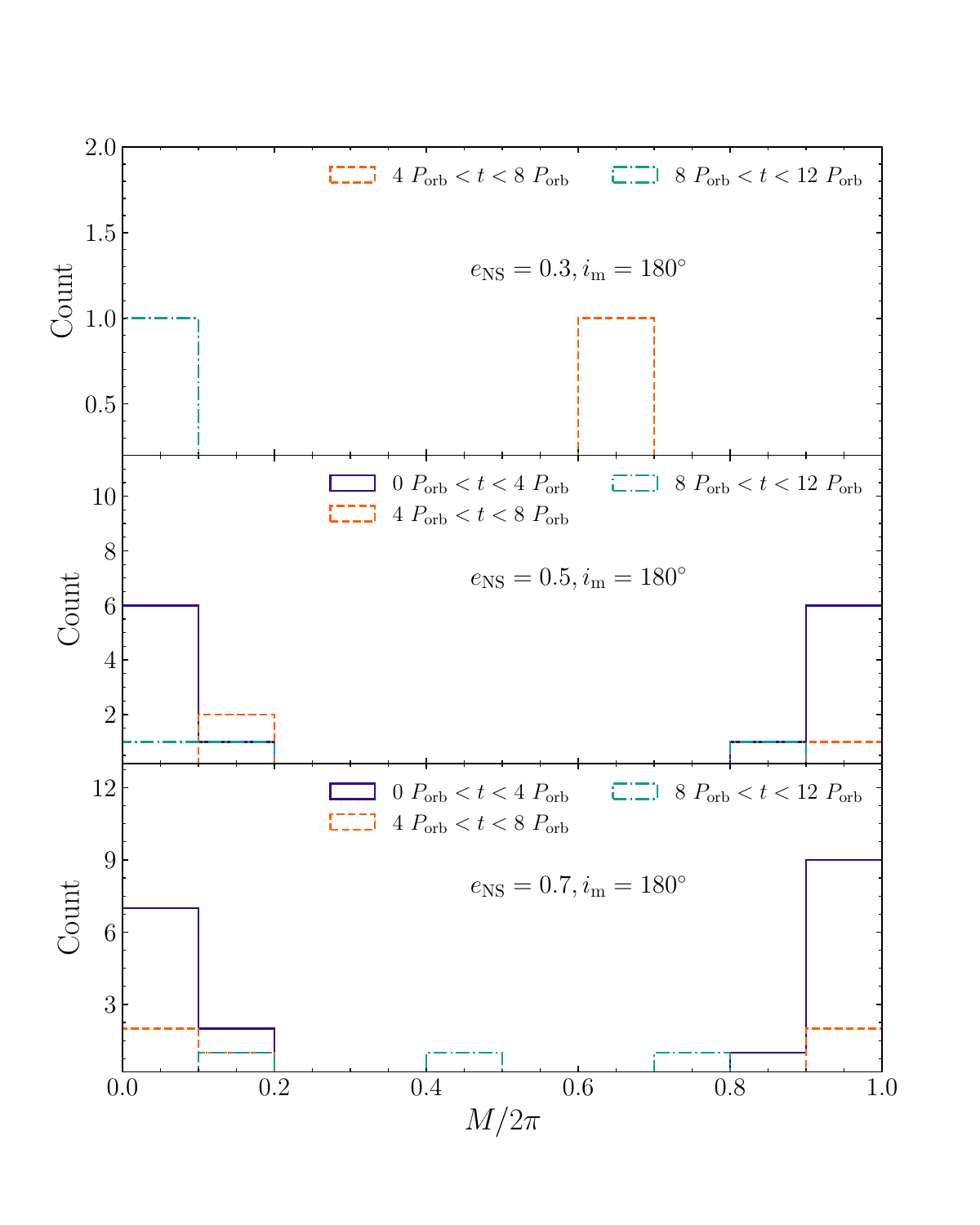}{0.49\textwidth}{}}
\caption{The evolution of the phase distribution of the collision
events for $i_{\rm m}=0^{\circ}$ (left panels) and $i_{\rm
m}=180^{\circ}$ cases (right panels) of Scenario \Rmnum{1}.
Note that in the case with $i_{\rm m}=180^{\circ}$ and
$e_{\rm NS}=0$, none of the particles collide with the neutron
star. In each panel, the solid line, dashed line, and dash-dotted
line represent the phase distribution during $0<t<4~P_{\rm orb}$,
$4~P_{\rm orb}<t<8~P_{\rm orb}$, and $8~P_{\rm orb} <t<12~P_{\rm
orb}$, respectively. The panels from top to bottom correspond to
$e_{\rm NS}$ = 0, 0.3, 0.5, and 0.7, respectively.
\label{fig10}}
\end{figure}

To conclude, we find that two conditions are good for producing
periodic collision events.  First, the neutron star orbit should
be nearly coplanar with respect to the main plane of the asteroid
belt, which will ensure a sufficient number of collision events.
Second, the orbit eccentricity of the neutron star should be
relatively large so that it can quickly traverse the asteroid belt.
In this way, the collisions will happen in a relatively narrow
phase window. For the coplanar cases (i.e. $i_{\rm m}=0^{\circ}$
or $180^{\circ}$), to further explore the periodicity and its
evolution, we have divided the first 12 neutron star orbital
periods into three stages, i.e. $0<t<4~P_{\rm orb}$,
$4~P_{\rm orb}<t<8~P_{\rm orb}$, and $8~P_{\rm orb}<t<12~P_{\rm orb}$.
The phase distribution of the collisions at each stage is shown
in Figure \ref{fig10}. In the case with  $i_{\rm m}=0^{\circ}$
and $e_{\rm NS}=0.3$, although the overall phase distribution
drawn from all the 300-period data shows a uniform feature,
we still notice that a large number of collisions actually happen
at around the periastron (i.e. $M/2\pi \sim 0$ and 1) at the early
state with $t<4P_{\rm orb}$. Further detailed examination reveals
that this is mainly attributed to the neutron star's first passage
through the asteroid belt. If the collision events of the first
period are expelled, the phase distribution of all other collisions
in the remained periods is highly uniform and no periodicity could
be observed.
Meanwhile, the collisions in the circular binary system
with $i_{\rm m}=0^{\circ}$ exhibit an approximately uniform
distribution even at the early stages (i.e. $t\leqslant4~P_{\rm
orb}$). In the case with  $i_{\rm m}=0^{\circ}$ and $e_{\rm
NS}=0.5$, we see that the  periodic behavior can be observed only
in the $0<t<4~P_{\rm orb}$ stage, and the periodicity disappears
when $t> 4 P_{\rm orb}$. Similarly, in the  the case with $i_{\rm
m}=0^{\circ}$ and $e_{\rm NS}=0.7$, the  periodicity exists only
when $t< 8 P_{\rm orb}$. In cases of $i_{\rm m}=180^{\circ}$, the
collisions mainly occur at the  $0<t<4~P_{\rm orb}$ stage and
centre on the phase of $M/2\pi$ = 0 or 1 (i.e. near the
periastron), hinting that the periodicity also exists. However,
the very limited collision number prevents us from drawing a firm
conclusion. A numerical simulation with a much denser asteroid
belt would help clarify this issue in the future.

\subsection{Scenario \Rmnum{2}: neutron star being on an internal trajectory}

\begin{figure}[ht!]
\gridline{\fig{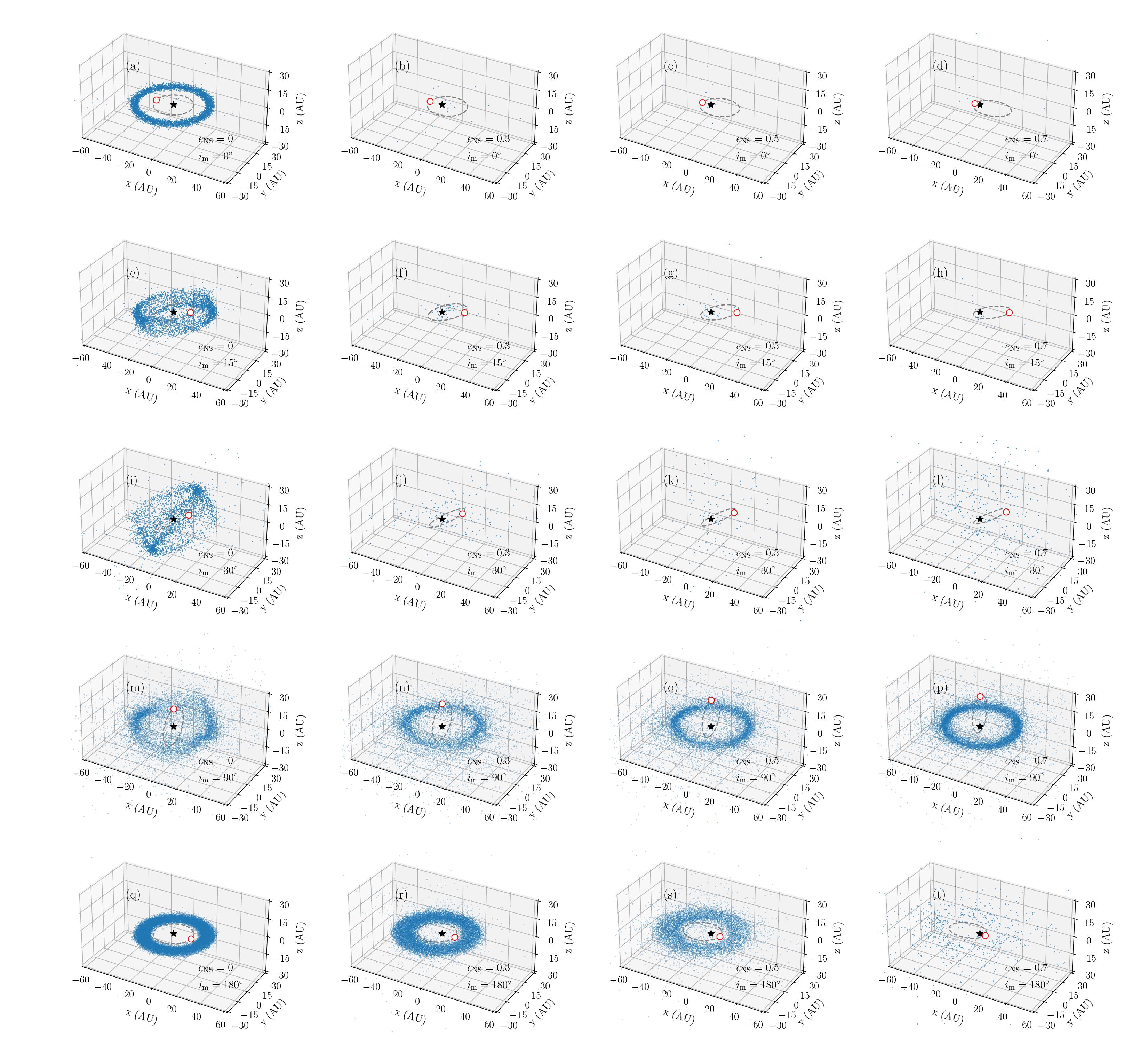}{1\textwidth}{}}
\caption{The spatial distribution of the asteroids after 300
periods of evolution (i.e. at the moment of $t=300~P_{\rm orb}$), for the
20 cases of Scenario \Rmnum{2}. Note that the initial conditions
are shown in Figure \ref{fig3} correspondingly. Line styles and symbols
are the same as those in Figure \ref{fig3}.
\label{fig11}}
\end{figure}

Figure \ref{fig11} shows the spatial distribution of the asteroids at
the final stage of our simulation (i.e. at $t=300~P_{\rm orb}$) for the
20 cases of Scenario \Rmnum{2} (whose initial distributions are illustrated
in Figure \ref{fig3}, correspondingly).
It is interesting to see that the asteroid belt is completely destroyed
in the cases with a non-zero $e_{\rm NS}$ and a low mutual
inclination (i.e. $i_{\rm m}\leqslant 30^{\circ}$). Only a small
fraction of asteroids are left, which are sparsely dispersed
throughout the 3D space.
 In Figure \ref{fig11} (e) and
(i), we see that the asteroids expands to the 3-D space in the
cases of $i_{\rm m}=15^{\circ}$ and $30^{\circ}$. This is due to
the precession of the angular momentum vector of the asteroids
when they orbit around the binary \citep{Chen2020MNRAS}. The node
also precesses in the circumbinary hydrodynamic disc, resulting in
either coplanar or polar alignment due to viscous dissipation
\citep{smallwood2023MNRAS}. In the cases with $i_{\rm
m}=90^{\circ}$ or $i_{\rm m}=180^{\circ}$, although the basic
structure of the asteroid belt is still preserved after
300 $P_{\rm orb}$ of evolution, the asteroids are
primarily distributed outside the neutron star orbit, which
indicates that their probability of collision with the neutron
star will be very low.

\begin{figure}[ht!]
\gridline{\fig{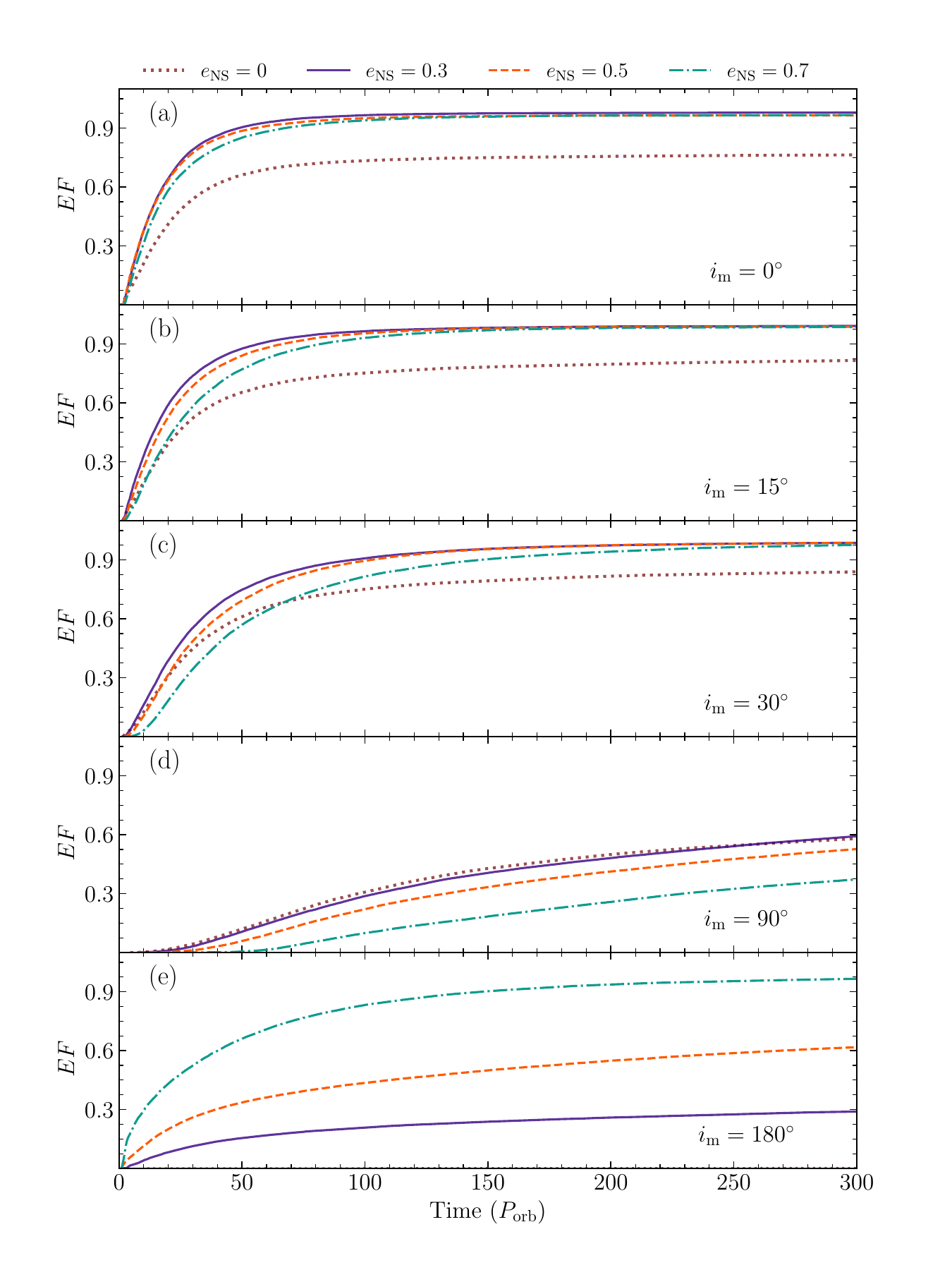}{0.8\textwidth}{}}
\caption{The cumulative ejection fraction ($EF$) of the asteroids
as a function of time for the 20 cases of Scenario \Rmnum{2}.
Time is measured in units of the neutron star's orbital period
$P_{\rm orb}$. The five panels from top to bottom correspond to
a mutual inclination of $i_{\rm m}=0^{\circ}$, $15^{\circ}$, $30^{\circ}$,
$90^{\circ}$, and $180^{\circ}$, respectively. The dotted line, solid
line, dashed line, dash-dotted line in each panel represent $e_{\rm NS}=$
0, 0.3, 0.5, and 0.7, respectively.
\label{fig12}}
\end{figure}

The temporal evolution of the $EF$ are plotted correspondingly in
Figure \ref{fig12}. In the cases where $i_{\rm m} \leqslant
30^{\circ}$ and $e_{\rm NS} \neq 0$, we have $EF \sim$ 0.9
at $t = $ 300 $P_{\rm orb}$, similar to the corresponding cases
in Scenario \Rmnum{1}. However, note that the neutron star
in a circular orbit exhibits a stronger scattering effect in
Scenario \Rmnum{2} than that of the corresponding cases in
Scenario \Rmnum{1}. When $i_{\rm m}=90^{\circ}$ and $i_{\rm
m}=180^{\circ}$, the scattering effect of the neutron star is
generally much lower as compared to other inclination cases.
We notice that the $EF$ continues to increase till
$t=300~P_{\rm orb}$ for the four cases in Figure \ref{fig12} (d).
Considering that these configurations do not effectively produce
collisions, the evolution of these systems is not tracked for
longer time. 

According to the stability plots of \cite{Chen2020MNRAS},
some configurations in Scenario \Rmnum{2} are unstable, especially
for those with $i_{\rm m} \leqslant 30^{\circ}$. However, the
setups and considerations are still meaningful. For example, a
neutron star in a low eccentricity retrograde orbit or in a high
eccentricity polar orbit cannot effectively scatter particles out
of the system, which will be relatively stable and could exist in
reality \citep{Chen2020MNRAS}. Additionally, for those unstable
configurations, although a portion of the asteroids will collide
with the binary or be ejected in a short period, the remaining
particles may still remain in the system for a long time, which
deserve detailed investigating as done in our simulations.

\begin{figure}[ht!]
\gridline{\fig{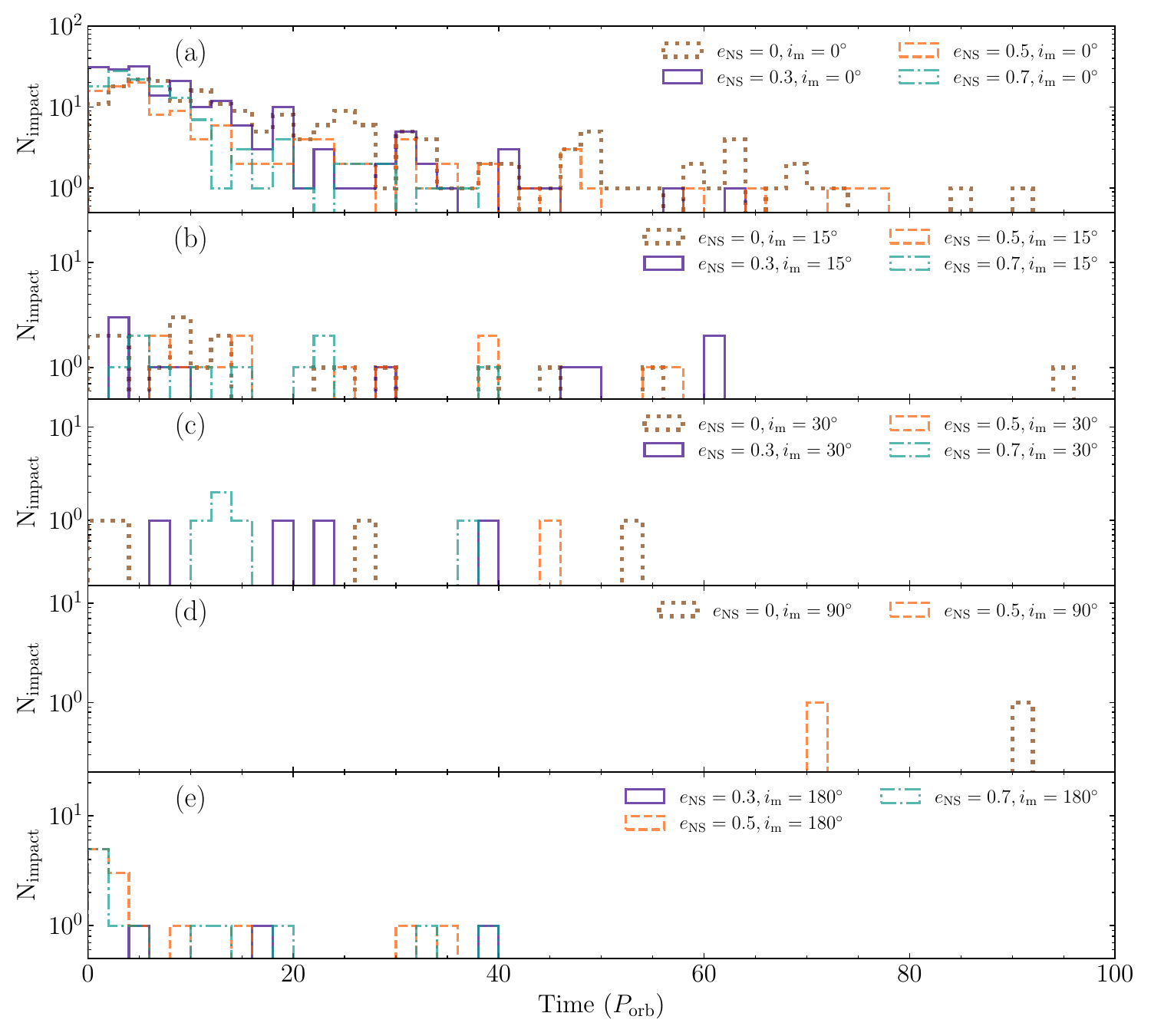}{0.9\textwidth}{}} \caption{The impact
number versus time for 17 cases of Scenario \Rmnum{2}.
The remaining 3 cases of Scenario \Rmnum{2} have no collision
events, thus are not shown in this plot. Time is measured in units
of the neutron star's orbital period $P_{\rm orb}$. The
five panels from top to bottom correspond to a mutual inclination
of $i_{\rm m}=0^{\circ}$, $15^{\circ}$, $30^{\circ}$,
$90^{\circ}$, and $180^{\circ}$, respectively. The dotted line,
solid line, dashed line, dash-dotted line in each panel represent
$e_{\rm NS}$ = 0, 0.3, 0.5, and 0.7, respectively.
\label{fig13}}
\end{figure}

The temporal evolution of the impact rate for Scenario \Rmnum{2} is
illustrated in Figure \ref{fig13}. It can be seen that the
$i_{\rm m}=0^{\circ}$ cases have a relatively high collision rate.
A lower orbital eccentricity of the neutron star results in an
increase in the collision rate, which is similar to the results
of $i_{\rm m}=0^{\circ}$ cases in Scenario \Rmnum{1}.
Additionally, the misaligned disk significantly
reduces the collision rate. Generally, the collision rate in
Scenario \Rmnum{2} is much smaller than that of Scenario \Rmnum{1},
the reason is that the majority of asteroids are beyond the range
of the neutron star orbit, which makes it extremely difficult for
the neutron star to capture the asteroids. To increase the collision
rate, we would need to significantly increase the number density
of asteroids in the belt, which would require a much higher
computational cost and is beyond the scope of this study.
Additionally, when the neutron star is on a retrograde orbit
(i.e. $i_{\rm m} = 180^{\circ}$), the collision rate is also
significantly reduced as compared to the $i_{\rm m}=0^{\circ}$ cases.

\begin{figure}[ht!]
\gridline{\fig{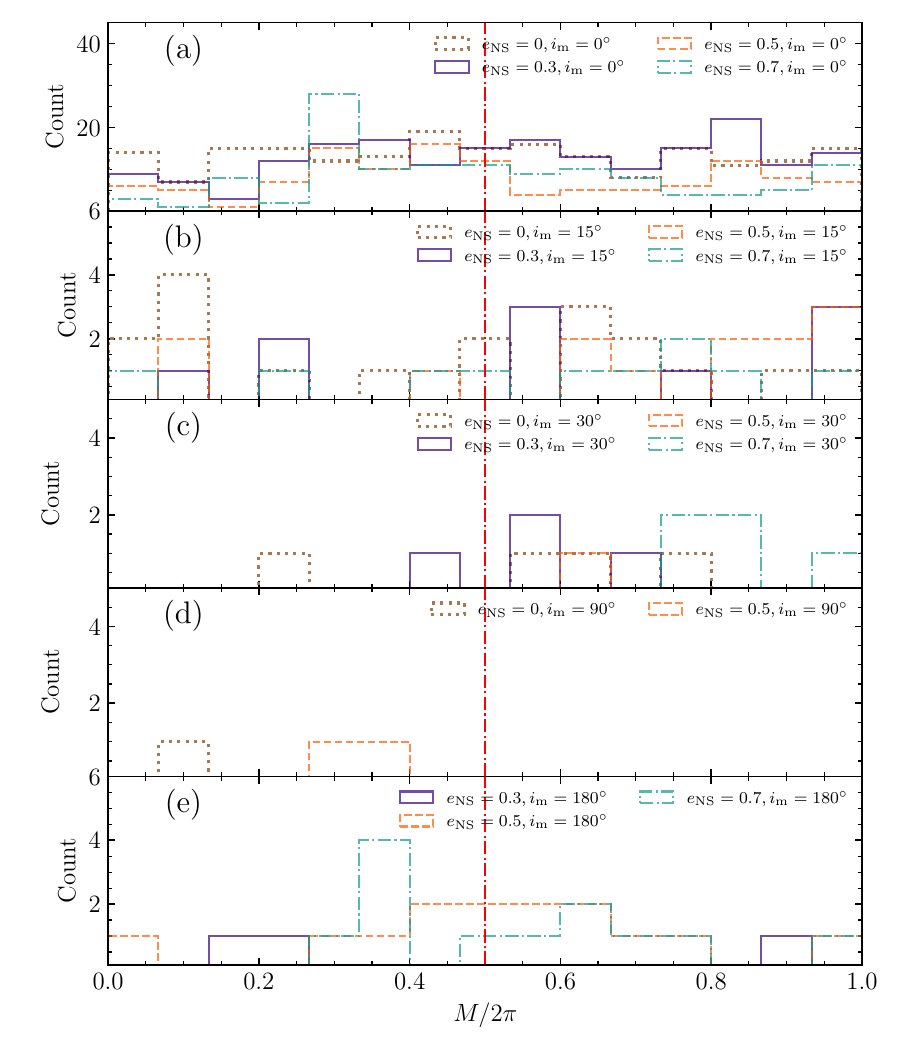}{0.85\textwidth}{}} \caption{The phase
distribution of the collision events when they are folded
according to the neutron star orbital period. Here the Y-axis is
the number of collision events and the X-axis is the mean anomaly
$M$ of the neutron star. Note that only 17 cases of
Scenario \Rmnum{2} are shown, since no collison events are
observed in the remaining 3 cases.
The five panels from top to
bottom correspond to the mutual inclination of $i_{\rm
m}=0^{\circ}$, $15^{\circ}$, $30^{\circ}$, $90^{\circ}$, and
$180^{\circ}$, respectively. $M = 0$ and $2\pi$ corresponds to the
periastron of the neutron star orbit, and the vertical dash-dotted
line at $M = \pi$ represents the apastron. The dotted line, solid
line, dashed line, dash-dotted line in each panel represent
$e_{\rm NS}$ = 0, 0.3, 0.5, and 0.7, respectively.
\label{fig14}}
\end{figure}

The phase distribution of the collision events when they are folded
according to the neutron star orbital period is shown
in Figure \ref{fig14} for Scenario \Rmnum{2}. We
see that the collision events generally do not cluster at any particular
orbital phase, thus no clear periodicity exists in the collisions.
Comparing with the coplanar cases of Scenario \Rmnum{1}, the neutron
star in Scenario \Rmnum{2} approaches the asteroid belt most closely when
it is at the apastron rather than the periastron. Hence, it is not difficult to
understand that the collisions are spreaded over a much broader phase range.
As a result, the cases in Scenario \Rmnum{2} are not good at producing
a periodic collision pattern.

\section{Conclusions and discussion}
\label{sec:summary}

The interaction between a neutron star and an asteroid belt are
investigated numerically in this study. According to the relative
position of the asteroid belt and the neutron star, the
configurations considered here can be grouped into two scenarios.
In Scenario \Rmnum{1}, the neutron star is assumed to be on an
external trajectory that intersects the asteroid belt; while in
Scenario \Rmnum{2}, the neutron star orbit is encompassed by the
asteroid belt. Various eccentricities and inclination angles are
assumed for the neutron star orbit. In Scenario \Rmnum{1}, when
the orbital eccentricity of the neutron star is high, the
periastron will penetrates further into the asteroid belt. As a
result, a greater number of asteroids will be ejected from the
system by the neutron star. Among all the configurations of
Scenario \Rmnum{1}, it is found that the case with $e_{\rm NS}=0.3$
and $i_{\rm m}=0^{\circ}$ exhibits the highest impact rate. However,
the collisions do not show any periodicity in time. When $e_{\rm NS}$
increases, the collisions tend to occur mainly near the periastron
so that they exhibit a periodic behavior to some extent. The most
obvious periodicity is observed in the case with $e_{\rm NS}=0.7$
and $i_{\rm m}=0^{\circ}$, for which the phase distribution of the
collisions deviates from the uniform distribution with a
$\chi^{2}_{\rm red}$ of $\sim$ 11.8. A detailed analysis
reveals that the periodic behavior of collisions can be sustained
for a time of $\sim 8 P_{\rm orb}$. For the cases of Scenario
\Rmnum{1}, it is also found that a low mutual inclination
(i.e. $i_{\rm m}\leqslant30^{\circ}$) can somewhat enhance
the stability of the system, leading to a slight reduction in the
ejection rate of the asteroids at early stages. However, the
collision rate is notably reduced since the asteroids are
scattered into the 3D space, which significantly decreases the
asteroid density along the orbital path of the neutron star.
Moreover, highly tilted systems (e.g. $i_{\rm
m}=90^{\circ}$) can lead to strong KL oscillations
\citep{smallwood2023MNRAS}, significantly increasing the number of
ejected asteroids and the portion of asteroids that collide with
the central massive star. In Scenario \Rmnum{2}, the
neutron star with a low eccentricity in a retrograde orbit or with
a high eccentricity in a polar orbit will scatter much less
asteroids out of the system, which is consistent with the
stability analysis of \cite{Chen2020MNRAS}. Generally, the
systems in Scenario \Rmnum{2} have a relatively lower collision
rate compared to the corresponding cases in Scenario \Rmnum{1}.
Additionally, no periodicity could be observed in the collision
events.

If FRBs are produced by the collision of asteroids with a neutron
star, then the lifespan of the FRB source will be linked to the
existence of the asteroid belt. Our study reveals that the initial
conditions of the system significantly affect the rate of the
ejections and collisions.
According to the stability analysis of \cite{Chen2020MNRAS},
most configurations in Scenarios \Rmnum{1} and \Rmnum{2} are
unstable, leading to the depletion of the asteroid belt after 300
$P_{\rm orb}$. Considering that a slight increase in the mutual
inclination (but with $i_{\rm m}<39^{\circ}$ to avoid KL
oscillations) can enhance the stability of the system, we cannot
rule out the possibility that a somewhat misaligned asteroid belt
could survive for a much longer time. Additionally, since the
asteroid belt is more or less destroyed to some extent every time
the neutron star passes it, leading to a continuous reduction in
the number density of asteroids, it is expected that the activity
of the FRB source would diminish as time elapses.


There are still many simplifications in our calculations. For
example, the asteroids are treated as test particles and their
gravitational force excerted on each other and on the neutron star
is omitted. The collision between two asteroids which may lead to
the generation of a large number of smaller clumps is also not
considered.
In fact, the number density and total number of
asteroids may be much higher than the setup in our simulations
(see \citealt{Smallwood2019MNRAS} for a detailed discussion). The
interaction between asteroids such as collisions or close
encounters may be non-negligible, potentially leading to more
complications in the process. Also, the distribution of the
asteroids in the belt could be highly inhomogeneous
\citep{Dai2016ApJ}. In realistic cases, the mass and radius of
individual asteroids should also vary from one object to another
\citep{Smallwood2019MNRAS}, which would lead to highly variable
FRB phenomena.  Finally, although we have explored many different
orbital configurations for the system, the actual parameter space
is far beyond our current study. More detailed numerical
simulations in the field are still necessary in the future.

\section*{acknowledgements}
We would like to thank the anonymous referee for useful
suggestions. This study was supported by the National Key R\&D
Program of China (2021YFA0718500),
the National Natural Science Foundation of China
(Grant Nos. 12233002, 12393812, 12041306, 11988101, U2031117, 12041303),
and the National SKA Program of China Nos. 2020SKA0120302 and 2020SKA0120200,
YFH also acknowledges the support from the Xinjiang Tianchi Program.
P.W. also acknowledges support from the CAS Youth Interdisciplinary Team,
the Youth Innovation Promotion Association CAS (id. 2021055), and the
Cultivation Project for FAST Scientific Payoff and Research Achievement
of CAMS-CAS.

\bibliography{sample631}{}
\bibliographystyle{aasjournal}

\end{document}